\documentclass[noeprint,twocolumn,superscriptaddress,amsmath,amssymb,aps]{revtex4-2}
\usepackage{graphicx}
\usepackage{float}
\usepackage{dcolumn}
\usepackage{bm}
\usepackage{physics}
\usepackage{xcolor}
\definecolor{darkblue}{rgb}{0,0,0.6}
\usepackage[colorlinks,linkcolor=darkblue,citecolor=darkblue,urlcolor=darkblue]{hyperref}
\usepackage[colorlinks,linkcolor=darkblue,citecolor=darkblue,urlcolor=darkblue]{hyperref}

\newcommand{\montpellier}{Laboratoire Charles Coulomb (L2C), Universit\'e de Montpellier, CNRS, 34095 Montpellier, France}

\newcommand{\cambridgeQ}{Yusuf Hamied Department of Chemistry, University of Cambridge, Lensfield Road, Cambridge CB2 1EW, United Kingdom}
\newcommand{\madison}{Department of Chemistry, University of Wisconsin–Madison, Madison, Wisconsin 53706, USA}

\begin{document}

\title{Front propagation in ultrastable glasses is dynamically heterogeneous}

\author{Cecilia Herrero}

\affiliation{\montpellier}

\author{Mark D. Ediger}

\affiliation{\madison}

\author{Ludovic Berthier}

\affiliation{\montpellier}

\affiliation{\cambridgeQ}

\date{\today}

\begin{abstract}
Upon heating, ultrastable glassy films transform into liquids via a propagating equilibration front, resembling the heterogeneous melting of crystals. A microscopic understanding of this robust phenomenology is however lacking because experimental resolution is limited. We simulate the heterogeneous transformation kinetics of ultrastable configurations prepared using the swap Monte Carlo algorithm, thus allowing direct comparison with experiments. We resolve the liquid-glass interface both in space and time as well as the underlying particle motion responsible for its propagation. We perform a detailed statistical analysis of the interface geometry and kinetics over a broad range of temperatures. We show that the dynamic heterogeneity of the bulk liquid is passed on to the front which propagates heterogeneously in space and intermittently in time. This observation allows us to relate the averaged front velocity to the equilibrium diffusion coefficient of the liquid. We suggest that an experimental characterisation of the interface geometry during the heterogeneous devitrification of ultrastable glassy films would provide direct experimental access to the long-sought characteristic lengthscale of dynamic heterogeneity in bulk supercooled liquids. 
\end{abstract}

\maketitle

\section{Introduction}

The physics of amorphous solids and the glass transition attracts considerable attention~\cite{berthier2016facets,debenedetti2001supercooled}. It offers both fundamental challenges to theoretical physicists who must deal with self-induced disordered structures relaxing in complex energy landscapes~\cite{berthier2011theoretical}, and experimental ones as characteristic lengthscales are often difficult to access over timescales which cover many orders of magnitude~\cite{ediger1996supercooled,angell2000relaxation}. In the last two decades these challenges were tackled in two major but relatively independent directions. First, a large amount of work was devoted to characterising the spatially heterogeneous dynamics of bulk supercooled liquids approaching the glass transition~\cite{ediger2000spatially,berthier2011dynamical}. Second, a novel family of amorphous materials was discovered when ultrastable glassy films were produced using physical vapor deposition in specific conditions~\cite{swallen2007}. Ultrastable glasses constitute a distinct class of amorphous solids with remarkable thermodynamic and kinetic stability, because they occupy deep minima in the energy landscape~\cite{ediger2017highly,rodriguez2022ultrastable,liu2015}.

Our main result is to reveal an unexpected connection between these two sets of questions. We have discovered that the dynamic heterogeneity of bulk supercooled liquids directly controls the heterogeneous devitrification of ultrastable glasses and affects the kinetics and geometric properties of the propagating front. We propose that devitrification via propagating fronts opens a unique path to the experimental determination of the characteristic lengthscale of dynamic heterogeneity in bulk supercooled liquids, whose direct determination remains an experimental challenge~\cite{tracht1998,reinsberg2001,berthier2005direct,dalle2007,zhang2018spatially}.

When a standard liquid-cooled glass is heated, the thermodynamically favoured liquid phase appears homogeneously within the glass matrix. By contrast, ultrastable films present such a tight molecular packing that the phase transformation is more easily initiated at the free surface, thus creating a liquid-glass interface that invades the glass at constant velocity~\cite{swallen2009,sepulveda2014,rodriguez2014,hocky2014,rodriguez2019,flenner2019,vlad2014}. Ultrastable glasses also transform differently in the absence of a free surface~\cite{kearns2010,vila-costa2020nucleation,fullerton2017density,herrero2022,ruiz2023,vila2023}. In both cases, the transformation of ultrastable glasses presents deep analogies with the melting of crystals.

The heterogeneous transformation via a propagating front has been extensively studied experimentally in several molecular glasses~\cite{sepulveda2012,rodriguez2014,rodriguez2015,tylinski2015,walters2015,rafols2017,dalal2015}. Experiments robustly report the existence of a sharp and uniform front separating the liquid from the glass, which propagates at constant velocity $v_f$ across distances which may exceed 1 micron. The velocity $v_f$ depends both on the stability of the glass and on the annealing temperature $T_a$ at which the transformation occurs \cite{vlad2014,walters2015,rafols2017}. For a given initial glass, the evolution of $v_f$ shows a clear correlation with the mobility of the bulk liquid at $T_a$. For instance the relation $v_f \sim \tau_\alpha^{-n}$, with $n \approx 0.6-1$ and $\tau_\alpha$ the structural relaxation time, has been reported in several works, extending across 10 orders of magnitude in $\tau_\alpha$  ~\cite{walters2015,rodriguez2015,rafols2017}. A connection with the diffusion constant $v_f \sim D^{n'}$ with $n' \approx 1-1.1$ has also been suggested~\cite{sepulveda2012}. This is reasonable as $D$ and $\tau_\alpha$ both quantify the dynamics of the bulk liquid, and are actually strongly correlated although slighlty decoupled from one another~\cite{stillinger1994translation,tarjus1995}. Remarkably, the correlation between $v_f$ and the bulk liquid dynamics appears to hold over a broad range of glass stabilities \cite{walters2015}. Stability in fact appears as a mere prefactor that decreases $v_f$ when the glass stability increases. Finally when the glass is annealed at high enough temperatures, a crossover between front propagation and bulk devitrification has been reported~\cite{whitaker2012,kearns2010,rodriguez2019}. In the very high temperature regime, recent experiments also suggest the existence of a surprising new regime where $v_f$ changes much faster with $T_a$ than either $D$ or $\tau_\alpha$~\cite{vlad2014,vlad2023}. 

Dynamic facilitation~\cite{fredrickson1984kinetic,chandler2010dynamics} provides a generic explanation for the existence of a propagating transformation front because the relaxing liquid near the interface can trigger the relaxation of the glass just across the interface~\cite{sepulveda2014manipulating}. As a result, front propagation can be theoretically analysed by constructing lattice models of dynamically facilitated glasses~\cite{leonard2010,gutierrez2016}, which reproduce the essential phenomenology described above. By introducing a form of kinetic facilitation in the framework of the random first order transition theory, predictions for the front propagation velocity were also obtained~\cite{peter2009spatiotemporal,wisitsorasak2013}. Computer simulations of atomistic models~\cite{berthier2023modern} have demonstrated the emergence of propagating fronts for increasingly stable glasses~\cite{lyubimov2013,hocky2014}. The recent development of the swap Monte Carlo algorithm has allowed the preparation of ultrastable glasses in computer models~\cite{equilibrium2016ozawa,ninarello2017,berthier2019swap}, whose transformation with a free surface was recently studied~\cite{flenner2019}. The relation $v_f \sim \tau_\alpha^{-1}$ was proposed~\cite{hocky2014,flenner2019}, but the dynamic range studied was limited to relatively high annealing temperatures. 

Due to the lack of molecular-level resolution in experiments, microscopic information on the front transformation kinetics and geometry are missing. For instance, although experiments and simulations have established that when the glass is annealed at a low enough temperature the front advances at constant velocity, insight on the geometry and roughness of the liquid-glass boundary and on its time evolution are yet to be established. In addition a microscopic picture of how particle motion in the supercooled liquid drives the propagation of the front should illuminate the link between $v_f$ and the dynamics of the bulk liquid, in order to rationalise the experimental findings and provide a solid microscopic picture of propagating fronts in ultrastable glasses. 

\begin{figure}
 \includegraphics[width=\linewidth]{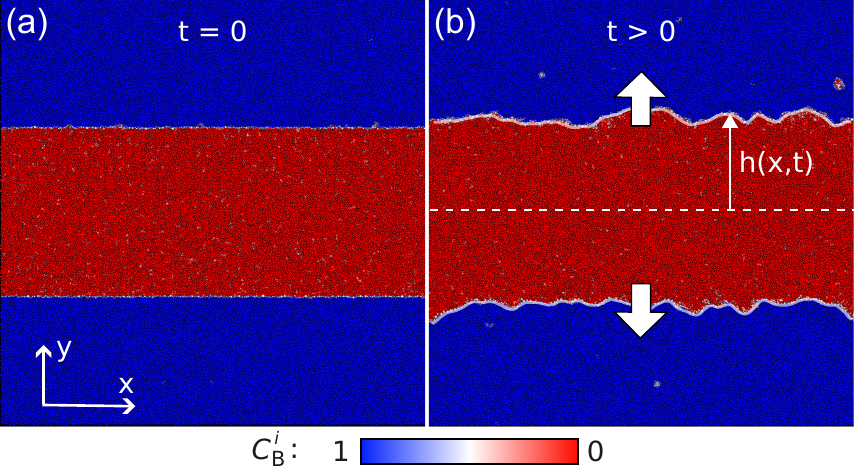}
 \caption{{\bf Simulation of a supercooled liquid front propagating into an ultrastable glass.}
   (a) Snapshot of the system, with $N=64000$ and $L = 253$, in its initial configuration, where an equilibrium liquid slit (red) is created inside an ultrastable glass matrix (blue).
   (b) As the liquid invades the glass region, the liquid-glass boundary $h(x,t)$ (white line) reaches a steady state where it moves at constant velocity $v_f$. Both snapshots correspond to $(T_i,T_a) = (0.035, 0.09)$. Particle colors reflect the value of the bond breaking correlation $C_B^i$ defined in Eq.~(\ref{eq:cb}).}
 \label{fig:fig1}
\end{figure}

We report molecular dynamics simulations of the transformation of ultrastable glasses via front propagation over a broad range of temperatures and glass stabilities. We use swap Monte Carlo to control the degree of stability of the initial states. We then introduce a liquid-glass boundary that we let evolve at constant annealing temperature $T_a$, see Fig.~\ref{fig:fig1}. This strategy allows us to resolve the position $h(x,t)$ of the liquid-glass interface in both space and time, to characterise its geometry with high resolution, and to relate its evolution to the underlying particle motion that we also resolve. 

For high annealing temperatures, the crossover between front and bulk mechanisms produces a complex transformation with an apparent acceleration of the front. At low temperatures, we measure a constant velocity $v_f$ over several decades in time with a temperature dependence in excellent agreement with experiments. In this regime, the front growth is heterogeneous and intermittent, because it is triggered by the spatially heterogeneous dynamics of the supercooled liquid. This observation allows us to relate $v_f$ to the diffusion constant $D$ of the liquid, which is decoupled from $\tau_\alpha$ at these low temperatures. 

Our results establish a deep connection between dynamic heterogeneity in bulk supercooled liquids and heterogeneous devitrification of ultrastable glassy films. We suggest that a careful experimental analysis of the geometry of propagating fronts using X-ray and neutron scattering techniques~\cite{xray1988sinha} will provide a direct experimental determination of the characteristic lengthscale of dynamic heterogeneity in supercooled liquids approaching the glass transition.

The manuscript is organised as follows. 
In Sec.~\ref{sec:model} we present the numerical model.
In Sec.~\ref{sec:observation} we describe how the front is initiated and characterise its average time evolution as well as its roughness. 
In Sec.~\ref{sec:evolution} we analyse the evolution of the averaged front velocity and its relation to bulk properties of the supercooled liquid. 
In Sec.~\ref{sec:growth} we resolve the space-time heterogeneity of the growth and its relation to the bulk relaxation mechanisms. 
We conclude the paper in Sec.~\ref{sec:conclusion} where we underlie the importance of an experimental characterisation of the front geometrical properties. 

\section{Numerical model and glass stability}

\label{sec:model}

We use molecular dynamics to simulate a two dimensional, size-polydisperse model of soft repulsive spheres, which is a well-characterized glass former~\cite{berthier2019zero,guiselin2022microscopic,thirty2022scalliet}. The pairwise interaction between particles $i$ and $j$, of diameters $\sigma_i$ and $\sigma_j$ and separated by a distance $r_{ij}$, is described by the potential:
\begin{equation}
    V_{ij}(r_{ij}) = \varepsilon \qty(\frac{\sigma_{ij}}{r_{ij}})^{12} + c_0 + c_2 \qty( \frac{r_{ij}}{\sigma_{ij}})^2 + c_4 \qty(\frac{r_{ij}}{\sigma_{ij}})^4,
\end{equation}
with $\sigma_{ij} = 0.5(\sigma_i + \sigma_j)(1 - \eta \abs{\sigma_i - \sigma_j})$. The parameters $c_0 = -28\varepsilon/r_{\rm c}^{12}$, $c_2=48 \varepsilon/r_{\rm c}^{14}$, and $c_4=-21\varepsilon/r_{\rm c}^{16}$, ensure the continuity of the potential up to its second derivative at the cutoff distance $r_{\rm c} = 1.25 \sigma_{ij}$. Size-polydispersity is introduced in the system through the probability distribution $\mathcal{P}(\sigma_i) = A \sigma_i^{-3}$, where $A$ is a normalization constant and $\sigma_i \in [\sigma_{\rm min},\sigma_{\rm max}]$, with $\sigma_{\rm min} = 0.73\sigma$, $\sigma_{\rm max} = 1.62\sigma$, and average diameter $\sigma$, which is used as the unit length. The non-additivity parameter is $\eta=0.2$. We use reduced units based on the particle mass $m$, the energy scale $\varepsilon$ and the average particle diameter $\sigma$, so the time unit is $\tau_0=\sigma \sqrt{m/\varepsilon}$. The timestep for numerical integration is $t_{\rm step} = 10^{-2} \tau_0$. We perform simulations in a square box of linear size $L$ with periodic boundary conditions.

In this study, we consider three different sets of initial conditions, corresponding to equilibrium states obtained using the swap Monte Carlo algorithm as described in Ref.~\cite{berthier2019swap} at temperatures $T_i = 0.035$, 0.06, 0.08 and constant number density $\rho = N/L^2 = 1$, corresponding to pressures $P_i = 2.31$, 2.66, and 2.94, respectively. For this density, the mode-coupling crossover temperature is $T_c \approx 0.12$, and the experimental glass transition temperature is at $T_g \approx 0.07$. Our three sets of initial conditions can therefore not be produced without the swap Monte Carlo algorithm. The lowest two temperatures $T_i$ are lower than $T_g$ and these initial conditions thus correspond to ultrastable glasses, in the experimental meaning of the word. 

\section{Observation of front propagation}

\label{sec:observation}

\subsection{Creation of liquid-glass fronts}

We create fronts between the supercooled liquid and the initial glass configuration and study its propagation within the $NPT$ ensemble. The methodology is as follows~\cite{hocky2014}. We start from an ultrastable glass configuration with $N=64000$ particles and number density $\rho=1$ (the linear box size is $L\approx 253$), equilibrated at a temperature $T_i$, and pressure $P_i$. 

First, we apply a Nos\'e-Hoover barostat to keep the pressure constant at $P_i$, and a thermostat to maintain the temperature at the desired annealing temperature $T_a > T_i$ for a short time $t=10^3$. This enables the initial configuration to rapidly expand and adjust to the new thermodynamic conditions $(P_i, T_a)$ while keeping the structure essentially unrelaxed.

Second, we prepare an interface by freezing the position of all particles outside a slit in the center of width $2 \ell_0$ with $\ell_0=50$. We have obtained equivalent results for $\ell_0=50$, 80 and 100. Particles in the center are transformed into a liquid by running $NVT$ simulations at very high temperature $T=0.8$ for a time $t=10^3$, which is sufficient to fully relax from the initial state in the central region. 

Third, we anneal the liquid region inside the cavity to the temperature $T_a$ using the swap Monte Carlo algorithm over a short period of duration $10^4$. 

Fourth, we simulate the glass to liquid transformation starting from this artificially-created liquid-glass flat interface by performing molecular dynamics simulations in the $NPT$ ensemble at conditions $(T_a, P_i)$ for the entire system. We have found similar results using the $NVT$ ensemble for the front kinetics, but the bulk kinetics can become unphysical if the $NVT$ ensemble is used~\cite{fullerton2017density,herrero2022}.

To increase the statistical significance of the results, we run five independent initial configurations for a given pair of temperatures $(T_i,T_a)$. As the liquid region is characterised by two independent interfaces, this gives a total of ten independent measurements for the front propagation. The error bars correspond to the statistical error at the $95\%$ confidence level.

\subsection{Tracking the position of the front}

Following previous work~\cite{herrero2022}, we use the bond-breaking correlation function $C_B^i(t)$ to distinguish liquid particles with $C_B^i \ge 0.5$, from glass particles with $C_B^i < 0.5$. We use this definition in the color code of Fig.~\ref{fig:fig1}, where red corresponds to liquid particles and blue to glass particles. 
The bond breaking correlation tracks changes in the local environment of a particle. It is defined as 
\begin{equation}
C_B^i(t)=\frac{n_i(t|0)}{n_i(0)},
\label{eq:cb}
\end{equation}
where $n_i(0)$ is the number of neighbours of particle $i$ at time $t=0$, and $n_i(t|0)$ the number of those particles that are still neighbours after time $t$. At $t=0$, neighbours are defined via the criterion $r_{ij}/\sigma_{ij}<1.3$. At $t>0$, we define neighbours via $r_{ij}/\sigma_{ij}<1.7$. These numbers were determined previously for this system~\cite{guiselin2022microscopic,thirty2022scalliet}. In equilibrium, the time decay of the average bond breaking correlation defines an equilibrium bulk relaxation time, $\langle C_B^i(t=\tau_B) \rangle = 1/e$, which we determine in a series of independent bulk simulations for each $(T_a,P_i)$.

To define the liquid-glass interface $h(x,t)$ shown in Fig.~\ref{fig:fig1}b, we discretize the particle coordinates on a lattice of square cells of linear size 1.47, chosen such that each cell contains at least one particle. Then, we binarize the average $C_B^i$ per cell, setting it equal to 0 (liquid) when $C_B^i \le 0.5$, or 1 when $C_B^i > 0.5$. The liquid-glass boundary is then identified with the liquid cells which are at the frontier with the glass matrix.

By construction, at $t=0$ the front between the liquid and glass regions is totally flat, see Fig.~\ref{fig:fig1}a. The liquid-glass interface $h(x,t)$ then freely evolves from its initial configuration for times $t>0$, as shown in Fig.~\ref{fig:fig1}b. The rest of the paper is mostly concerned with a detailed characterisation of the statistical properties of the front lines $h(x,t)$, and we borrow observables and vocabulary from the field of surface growth analysis~\cite{barabasi1995fractal,meakin1998fractals}.  

\subsection{Averaged evolution of the front}

The main observation from Fig.~\ref{fig:fig1}b is that for sufficiently stable initial configurations the liquid-glass front created artificially at $t=0$ is maintained at $t>0$ as the supercooled liquid slowly invades the ultrastable glass matrix. In other words, our simulations are able to reproduce the heterogeneous transformation process of ultrastable glasses reported experimentally, which represents a numerical accomplishment in itself~\cite{berthier2023modern}. 

To characterize the kinetics of this front propagation, we first define the spatially averaged front position at time $t$, 
\begin{equation}
\langle h \rangle (t) = \frac{1}{L} \int_0^L dx' h(x',t),
\end{equation}
where the brackets implicitly imply an average over independent realisations. 

\begin{figure}
 \includegraphics[width=\linewidth]{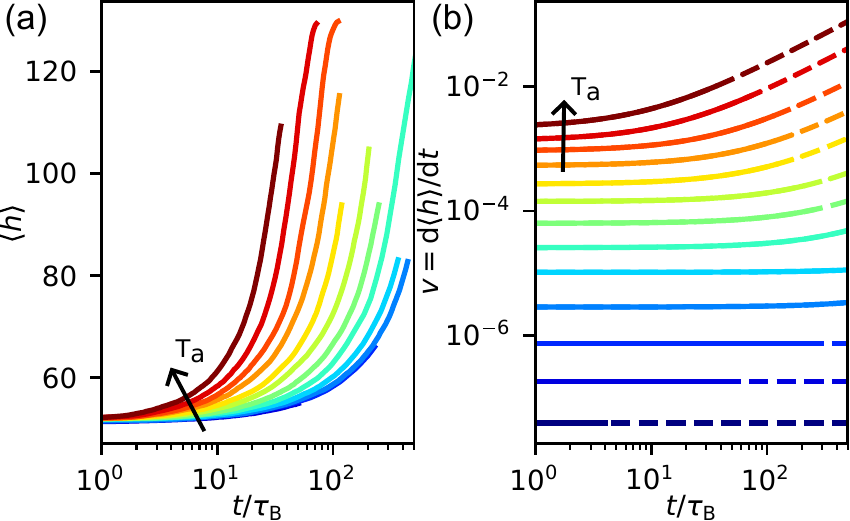}
 \caption{{\bf Spatially-averaged evolution of the front.}
   (a) Time evolution of the average front position $\expval{h}(t)$ for $T_i = 0.035$ and $T_a \in [0.08,0.14]$ increasing from bottom to top. The time is normalized by the bulk relaxation time $\tau_B$ of the liquid.
   (b) Time evolution of the instantaneous front velocity defined as the time derivative of the average interface position, rescaled by $\tau_B$. Full lines represent the time domain over which we collect data, dashed lines are extrapolations of the fitting function.}
 \label{fig:fig2}
\end{figure}

In Fig.~\ref{fig:fig2}a we show results for $\expval{h}(t)$ for the most stable glass we analysed with $T_i = 0.035$, and different annealing temperatures $T_a \in [ 0.08, 0.14]$. Anticipating that the timescale of the front propagation is mostly controlled by the bulk relaxation time of the liquid at $T_a$, we normalized $t$ by the characteristic time of the bulk liquid, $\tau_B(T_a)$.   

When analysing the evolution of $\langle h \rangle (t)$, one must be careful about the time domain considered. At very short times, the front geometry evolves from a flat to a more rough geometry before reaching a dynamic steady state during which its statistical properties become invariant with respect to time. However, at very long times the two fronts shown in Fig.~\ref{fig:fig1} may begin to interact due to the periodic boundary conditions in a finite system. This effect can be noticed in Fig.~\ref{fig:fig2}a at long times where $\langle h \rangle$ seems to saturate for the highest $T_a$ values.  

One can see in Fig.~\ref{fig:fig2}a that the front propagates faster for higher $T_a$. To quantify these effects, we fit our data with a second-order polynomial of the form: $\expval{h}(t) = a + bt + ct^2$ for the restricted interval $\expval{h} \le 100$ in order to avoid the effect of interactions with the periodic images at longer times. 

From this fit we can define the instantaneous front velocity, \begin{equation}
    v(t) = \frac{ d \expval{h} }{dt},
\end{equation}
determined numerically as $v(t) = b + 2ct$ and shown in Fig.~\ref{fig:fig2}b. Here, one observes that a small change in $T_a$ can produce orders of magnitude differences in the velocity of the front propagation. We shall analyse the temperature evolution of these data in the following.

From the above analysis, we can also define from the expression of the instantaneous velocity a characteristic timescale $t^* = b/c$, such that for times $t<t^*$ the front propagation velocity is essentially constant with value $v_f = b$. From Fig.~\ref{fig:fig2}b, we see that $t^*$ enters our time window when $T_a$ increases, and it becomes increasingly shorter at higher $T_a$ values. In this high temperature regime, the propagation of the front does not occur at a constant velocity and the front in fact accelerates as the transformation proceeds. For these temperatures, we again define the velocity as $v_f=b$, but note that its determination becomes ambiguous as it is limited to a small time window where it is not clear that steady state has been reached.  

\subsection{How rough is the propagating front?}

As the liquid-glass front moves from its initial position, its roughness also increases, as seen in Fig.~\ref{fig:fig1}. When a constant value of the velocity is reached, a dynamic steady state is reached in which two scenarios are possible. Either the interface develops fluctuations on increasingly larger lengthscales and its variance would keep increasing without bound, or the fluctuations saturate at a given scale to maintain a constant roughness. Such scenarios are typically observed in different models for interface growth~\cite{barabasi1995fractal,meakin1998fractals}.   

\begin{figure}
 \includegraphics[width=\linewidth]{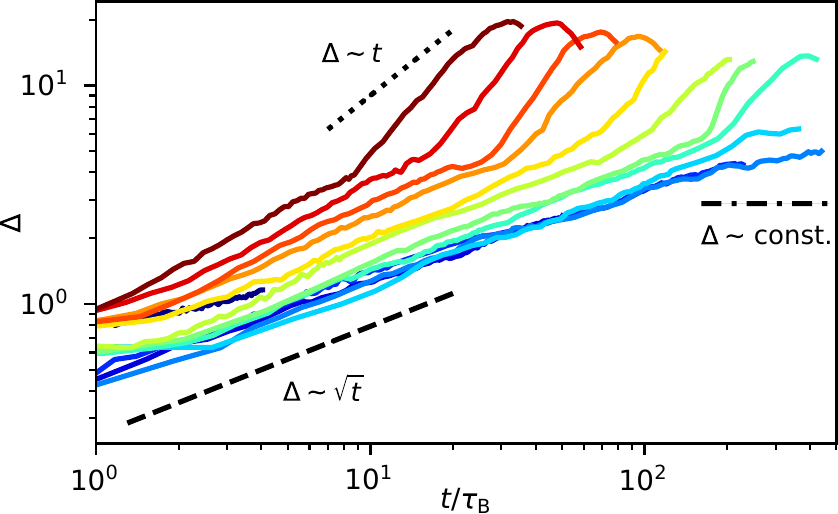}
 \caption{{\bf Evolution of the front roughness.}
   Time evolution of the front roughness characterised by the standard deviation $\Delta(t)$ in Eq.~(\ref{eq:standard}) for $T_i = 0.035$ and $T_a \in [0.08,0.14]$. The time is normalized by the bulk relaxation time $\tau_B$ of the liquid. The roughness grows diffusively at short times, but the front broadens rapidly ($\Delta \sim t$) at high temperatures when it accelerates. In contrast, the roughness saturates at long times at low temperatures, $\Delta \sim const$.}
 \label{fig:fig3}
\end{figure}

To quantitatively follow the evolution of the front roughness, we first define the second moment of the interface position  
\begin{equation}
    \langle h^2 \rangle (t) =  \frac{1}{L} \int_0^L dx' h^2(x',t),
\end{equation}
from which we can define the standard deviation 
\begin{equation}
    \Delta(t) = \sqrt{\langle h^2 \rangle - \langle h \rangle^2}.
\label{eq:standard}
\end{equation}
Clearly, $\Delta(t)$ physically represents the typical excursion of the interface away from its average position.  

In Fig.~\ref{fig:fig3}, we present numerical results for the time evolution of $\Delta(t)$ at different temperatures $T_a$. Several regimes can be observed. For all temperatures, the standard deviation $\Delta(t)$ grows from a very small value corresponding to the initial flat interface, and we find that it grows diffusively in this regime, $\Delta(t) \sim \sqrt{t}$, with only a weak temperature dependence. 

The behaviour of $\Delta(t)$ at long times depends sensitively on the annealing temperature $T_a$. For high temperatures, we noted earlier that the front accelerates after a typical timescale $t^*$. Visual inspection shows that this happens when the front encounters liquid domains that are already formed in the bulk ahead of it. When a liquid droplet merges with the front, its shape is suddenly affected, its roughness increases, and the front broadens. This crossover is very clearly observed in the standard deviation in Fig.~\ref{fig:fig3}, and it leads to a nearly-linear growth of the roughness, $\Delta(t)\sim t$. This very fast linear scaling presumably results from the fact that larger and larger droplets are absorbed as the front keeps moving, since the bulk keeps transforming. 

At low temperatures instead, when the velocity remains constant at large times, a dynamic steady state is reached where the roughness of the interface appears to saturate to a finite value, see Fig.~\ref{fig:fig3}. This regime is difficult to access numerically, as very large times (in units of the large bulk relaxation time $\tau_B$ itself) are needed, but this regime is only observed at very low temperatures where bulk melting does not interfere with the propagating front. We have performed careful visual inspections of individual trajectories, to confirm that it is the merging of small liquid droplets formed in the bulk which is responsible with the increase of the roughness at very long times.  

A diffusively growing standard deviation is easily explained by assuming independent relaxation events happening at distributed times $t$ at different locations $x$~\cite{tylinski2015,meakin1998fractals} in which case the roughness grows with no bound. Our finding that the roughness saturates at a finite value of order $\Delta \sim 10$ at large times and low temperatures suggest the existence of a physical mechanism that prevents the interface from becoming rough. In Sec.~\ref{sec:growth} we will argue that dynamic facilitation provides such a mechanism and we will provide microscopic support from our simulations leading to the physical picture of a relatively flat interface moving at constant velocity in a dynamic steady state. 

\section{Evolution of front propagation velocity}

\label{sec:evolution}

\subsection{Evolution with temperature and glass stability} 

\begin{figure}
 \includegraphics[width=\linewidth]{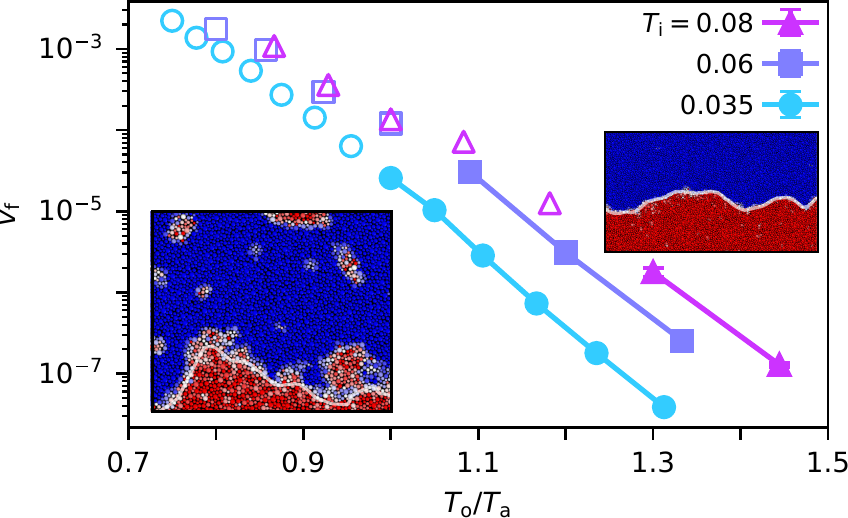}
 \caption{{\bf Evolution of the front velocity.}
    Temperature dependence of the front velocity $v_f$ for three glasses with different stabilities characterised by $T_i$. The temperature axis is normalized by the onset temperature $T_o$. Full points correspond to temperatures with constant front growth and well-defined front [right snapshot for $(T_i,T_a)=(0.035,0.09)$],
    empty symbols when an acceleration is observed due to appearance of liquid droplets in the bulk of the glass [left snapshot for $(T_i,T_a)=(0.035,0.14)$]. Color code of snapshots as in Fig.~\ref{fig:fig1}. Errorbars are of the order of the symbol size.}
 \label{fig:fig4}
\end{figure}

We show in Fig.~\ref{fig:fig4} the temperature dependence of the front velocity, using the rescaled variable $T_o/T_a$ dependence with $T_o$ marking the onset temperature for slow dynamics. The onset differs for the three different glasses considered, as they are studied at different pressures $P_i$. Therefore, rescaling the annealing temperature by $T_o$ trivially removes the effect of studying different pressures. We use open symbols for the temperatures at which an acceleration of the velocity is observed and empty symbols when $v_f$ can be more robustly defined at low temperatures. This distinction allows us to distinguish two distinct regimes: one at high temperatures (empty symbols) where the glass stability seems to play a negligible role, and another at low temperatures (full symbols) where the velocities differ more drastically for different glass stabilities. 

Empty symbols at high $T_a$ in Fig.~\ref{fig:fig4} correspond to the temperatures where acceleration is observed in the instantaneous velocity plots (as in Fig.~\ref{fig:fig2}b). Deviations from constant growth have been previously reported in experiments and related to a significant broadening of the glass-liquid boundary~\cite{vlad2023}. However, due to the lack of microscopic resolution, this hypothesis could not be verified in the experiments. We can confirm this scenario from our simulation results. We show in Fig.~\ref{fig:fig4}(left) a snapshot of the liquid-glass frontier for $(T_i,T_a) = (0.035, 0.14)$, representative of the high $T_a$ behavior. At these high temperatures, we observe that the glass starts to transform also in the bulk, which takes the form of rare droplets of liquid which slowly nucleate~\cite{herrero2022}. When the propagating front hits one of these droplets, its position suddenly advances by a large amount. As a result the progression of the front accelerates. As time increases, more and more droplets are encountered and the velocity gets larger. The acceleration of the apparent front velocity arises because the front has travelled a distance comparable to the crossover length separating bulk from front melting~\cite{kearns2010}, and bulk melting starts to significantly influence the front propagation.    

By contrast, when the crossover length is much larger than the distance travelled by the front, as happens at low $T_a$, the front velocity is well-defined and remains constant over several decades of times, see Fig.~\ref{fig:fig2}b. In this regime, the liquid-glass front is very well-defined (see the right snapshot in Fig.~\ref{fig:fig4}), as no relaxation dynamics takes place in the untransformed glass matrix ahead of the front. The data in Fig.~\ref{fig:fig4} indicate a very strong evolution with $T_a$ for a given glass stability, which is either Arrhenius or slightly super-Arrhenius, as discussed below. When changing the stability of the glass by varying $T_i$, we observe that the $T_a$-dependence remains essentially the same, with stability entering as a simple prefactor which varies by about $20$ between our different glasses. These results are equivalent to previous experimental observations~\cite{rodriguez2015,walters2015} and simulations~\cite{flenner2019}. It is remarkable that a large change of glass stability can produce such a large variation of the front velocity at low temperatures. This result is again in close agreement with experiments~\cite{walters2015,rafols2017}.

\subsection{Relation to equilibrium bulk properties}

Following experimental investigations, we now compare the velocity measured in the situation of front propagation to various dynamic properties of the corresponding supercooled liquid at temperature $T_a$. The rationale for such a comparison is quite simple: the liquid-glass interface that moves separates a nearly equilibrated supercooled liquid with well-defined dynamic properties that are uniquely controlled by $T_a$, from an ultrastable glass for which no characteristic timescale for relaxation exists (or the timescale is so large that it cannot be measured~\cite{how2020berthier}). 

In the low-$T_a$ regime that we wish to understand, structural relaxation only occurs in the liquid region, which slowly erodes the stable glass matrix at a constant velocity. We have determined two characteristic timescales for the bulk liquid. We measured both the self-intermediate scattering function $F_s(q,t)$ and the mean-squared displacement $\Delta_s^2(t)$ derived from the self-motion of the particles. Because our simulations are performed in two dimensions, we follow common practice and use the cage-relative coordinates to estimate these correlators~\cite{illing2017}. We then obtain the structural relaxation time $\tau_\alpha$ from  $F_s(q,t=\tau_\alpha) = 1/e$, using $q=6.9$. The diffusion coefficient $D$ is obtained from the long-time limit of $\Delta_s^2(t) \sim 4 D t$, as usual (more generally $\Delta_s^2 \sim 2d Dt$ in $d$ dimensions).    

\begin{figure}
 \centering
 \includegraphics[width=\linewidth]{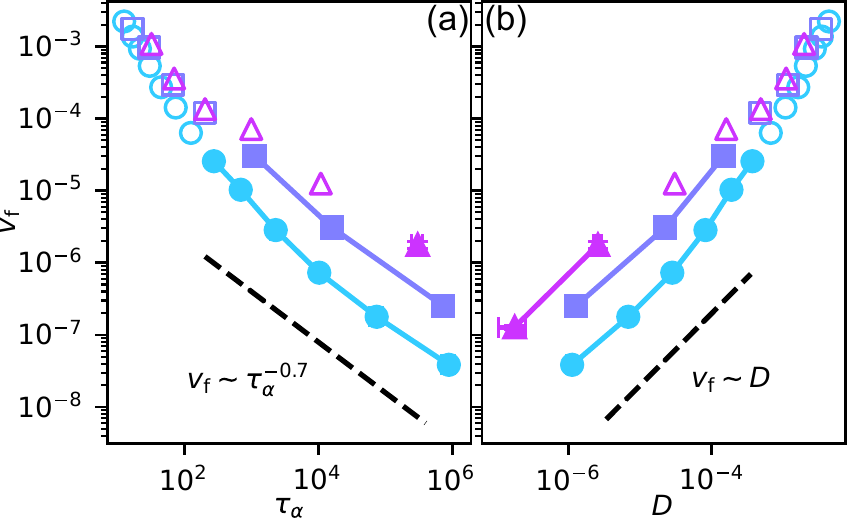}
 \caption{{\bf Correlation between front velocity and equilibrium bulk dynamics.}
 Front velocity $v_f$ in glasses of different stabilities as a function of 
 (a) the relaxation time $\tau_\alpha$ defined from the self-intermediate scattering function, and 
 (b) the self-diffusion constant $D$ of the bulk liquid at $T_a$.
Dashed lines are guides indicating (a) $v_f \sim \tau_\alpha^{-0.7}$ and $v_f \sim D$. Symbols and colors as in Fig.~\ref{fig:fig4} with errorbars of the order of the symbol size.}
\label{fig:fig5}
\end{figure}

First, as previously done in many experiments~\cite{sepulveda2012,rodriguez2015,walters2015} and simulations~\cite{hocky2014,flenner2019}, we study the correlation between $v_f$ and $\tau_\alpha$ in Fig.~\ref{fig:fig5}a. Focusing on the low temperature regime where the velocity $v_f$ is well-defined, we observe a strong correlation between the two quantities, as expected, but quantitatively we observe a power law relation, $v_f \sim \tau_\alpha^{-n}$ with an exponent $n \approx 0.7$. This result is in good agreement with previous experimental work, where the results are systematically consistent with an exponent $n<1$~\cite{sepulveda2012,rodriguez2015,walters2015}.

We then compare the evolution of $v_f$ to the one of the diffusion constant $D$ in Fig.~\ref{fig:fig5}b. Here again the correlation is strong, but the interrelation between the two quantities appears much simpler, since at low temperatures we find a direct proportionality relation between them, namely 
\begin{equation}
v_f \sim D. 
\label{eq:vD}
\end{equation}
Direct comparisons between $D$ and $v_f$ are scarce in the experimental literature, as $D$ is more difficult to measure than $\tau_\alpha$ but the results in Ref.~\cite{sepulveda2012} are indeed compatible with Eq.~(\ref{eq:vD}). Being able to reach low enough temperatures and a broad enough dynamic range to establish the validity of Eq.~(\ref{eq:vD}) is an important achievement of the present work.   

In both panels of Fig.~\ref{fig:fig5}, we observe a change of the evolution of the velocity at high $T_a$ (corresponding to empty symbols as in Fig.~\ref{fig:fig2}a). This apparent change was also reported in previous experiments~\cite{vlad2014} and simulations~\cite{flenner2019}. Analogously to the results described in Fig.~\ref{fig:fig4}, we relate this change of behavior to the crossover between bulk devitrification and front transformation: due to the presence of liquid droplets in the transforming bulk glass which affect the front propagation, the front velocity is no longer exclusively determined by bulk liquid properties as bulk nucleation interferes with the front growth, and its value is anyway difficult to precisely measure due to the absence of a well-defined steady state regime.

\section{Heterogeneous front growth}

\label{sec:growth}

\subsection{Characterizing front growth heterogeneity}

To get a deeper understanding of the front propagation kinetics, we now focus on spatio-temporal fluctuations of the growth relative to the average behaviour. Having access to the front position $h(x,t)$ at all times with atomistic resolution allows us to follow the complete history of the propagating front in space and time. 

\begin{figure}
 \centering
 \includegraphics[width=\linewidth]{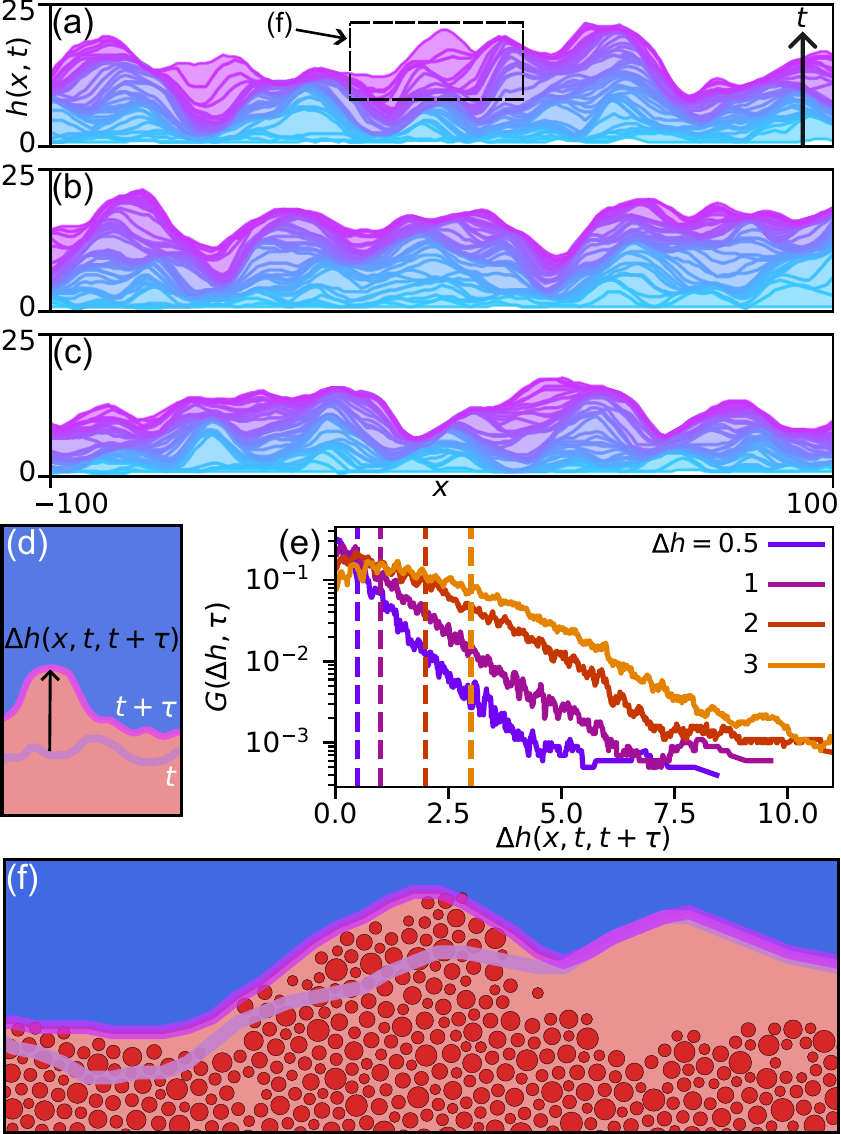}
 \caption{{\bf Front growth is dynamically heterogeneous.}
   (a-c) Space-time representation of the front for three independent initial glass configurations with $(T_i,T_a)=(0.035,0.09)$. The interface is shown every $\tau = 0.5/v_f$, with time growing from blue to purple. The aspect ratio is 1:1.5, thus accentuating the interface roughness.
   (d) Schematic representation of the local growth $\Delta h (x,t,t+\tau)$.
   (e) Probability distribution function of $\Delta h$ for times $\tau v_f = 0.5$, 1, 2, and 3 (from left to right). Dashed lines indicate the first moment of the distribution.
   (f) Inset of (a) highlighting that local relaxation in the fluid triggers front propagation. We draw the front position at two times separated by $\tau$, as well as the fluid particles which were mobile over the same time interval (red). The front does not propagate where the liquid is immobile (particles not shown). The size of the box is $40 \times 10$ (with aspect ratio 1:1).}
 \label{fig:fig6}
\end{figure}

In Figs.~\ref{fig:fig6}a-c, we show the time evolution of the front $h(x,t)$ for three independent realisations with the same parameters $(T_i,T_a)=(0.035,0.09)$.  In these images, we represent the position of the interface at different times separated by an interval $\tau$ such that the averaged distance travelled by the front is $0.5 \sigma$, i.e. $\tau = 0.5 / v_f$. This timescale is chosen to be large enough to observe significant displacements between each time lapse, and yet small enough to inform us about the detailed history and heterogeneity of the front displacement.  

We see in Figs.~\ref{fig:fig6}a-c that, at short times, front propagation rapidly deviates from the manually created flat interface shown in Fig.~\ref{fig:fig1}a. This flat interface develops spatial heterogeneities, with some regions growing much faster than others. As time increases, all regions eventually move, and the interface reaches a dynamic steady state with a small but finite roughness and a constant averaged velocity (see Fig.~\ref{fig:fig2}). These time series primarily demonstrate that the front does not propagate homogeneously in space, and that at any given location the propagation is very intermittent in time: some regions which appear nearly immobile over long periods of time finally move very fast at some later period. Overall, these images capture the main insights gained from the molecular dynamics simulations as they directly reveal the spatially heterogeneous and intermittent nature of the front propagation kinetics.

The study of spatially heterogeneous and intermittent dynamics in bulk supercooled liquids has a long history~\cite{ediger2000spatially,berthier2011dynamical} from which we can directly borrow tools to quantify the heterogeneous front propagation. In bulk liquids, dynamic heterogeneity can be captured by the distribution of single particle displacements, also called the van Hove distribution~\cite{kob1997dynamical,universal2007chaudhuri}. We can define a similar quantity for propagating fronts. To this end, we first introduce the local displacement of the front position at a given position $x$ over an interval $\tau$ as
\begin{equation}
    \Delta h(x; t, t+\tau) = h(x,t+\tau) - h(x,t),
\end{equation}
as illustrated in Figure~\ref{fig:fig6}d. We then introduce the corresponding probability distribution function 
\begin{equation}
G (\Delta h, \tau) = \langle \delta \qty(\Delta h - \Delta h(x,t,t+\tau) ) \rangle,
\end{equation}
where the average is taken over different realisations, the positions $x$, and various times $t$ all taken in the steady state, so that the distribution $G(\Delta h,\tau)$ in effect only depends on $\tau$. If the front propagation were homogeneous in space, one would expect the probability distribution to be sharply peaked around an average value given by $\Delta h = v_f \tau$. 

The data in Fig.~\ref{fig:fig6}e instead reveal broad probability distributions for different $\tau$ values corresponding to averaged travelled distances between 0.5 and 3 indicated by vertical dashed lines. This confirms the impression provided by the snapshots in Figs.~\ref{fig:fig6}a-c that some regions can propagate several molecular diameters more than the average while others are nearly immobile. For instance for $\tau = 1/ v_f$ and an average distance of 1 particle diameter, we observe that some rare regions of the interface can move up to 7 times more than the mean, and that over the same period a large majority of the statistical weight corresponds to extremely small displacements. In other words, in time frames separated by roughly one relaxation time of the bulk liquid, most parts of the interface are actually immobile while some rare regions move quite large distances. This qualitative description echoes the analogous description of dynamic heterogeneity in bulk supercooled liquids. This analogy is made stronger by the observation of a roughly exponential decay of the the distribution $G(\Delta h, \tau)$ in Fig.~\ref{fig:fig6}(e), similarly to ubiquitous observations in bulk glassy liquids~\cite{universal2007chaudhuri}.  

These results demonstrate that the front propagation is dynamically heterogeneous in a way that is very similar to the bulk dynamics of the equilibrium liquid itself. This is physically natural, as the only particle motion and microscopic relaxation events that are responsible for the propagation of the front actually occur in the supercooled liquid side of the front, which is known to be dynamically heterogeneous at sufficiently low $T_a$.

\subsection{Relation to bulk heterogeneous dynamics}

To make the connection between front propagation and molecular motion in the liquid clearer, we show in Fig.~\ref{fig:fig6}f a zoom of the liquid-glass boundary taken from Fig.~\ref{fig:fig6}a, showing the front position at two different times separated by $\tau = 0.5 / v_f$, together with the particles that have relaxed (i.e. lost more than half of their neighbors) between these two times. We clearly observe that the regions where the front propagates significantly over a given time interval also correspond to those where the liquid has also relaxed. This implies that the spatially heterogeneous front propagation is in fact a direct consequence of the underlying spatially heterogeneous relaxation of the supercooled liquid. Consistently, the regions that did not grow at all between $t$ and $t+\tau$ correspond to regions of the bulk liquid where structural relaxation did not occur in this particular time period.

In this view, a relaxing domain within the supercooled liquid is able to push the front by a distance $\Delta h(x,\tau)$ which is larger than the average. Crucially, another consequence is that the spatial fluctuations of the front line along the direction $x$ should  be dictated by the typical size of the relaxing domains in the bulk liquid, i.e. to the bulk dynamic correlation length, which is often called $\xi_{4}$~\cite{berthier2011dynamical}.

To confirm this intuition, we define the equal-time spatial correlation of the height $h(x,t)$ in the longitudinal direction as 
\begin{equation}
  C(r) = \langle  h(x,t) h(x+r,t)  \rangle - \langle h \rangle^2 (t) . 
\label{eq:Cr}
\end{equation}
When measured in steady state at low temperatures, this function only depends on $r$ but not on the time $t$, by construction. By definition also, $C(r)$ is related for $r=0$ to the roughness defined in Eq.~(\ref{eq:standard}), since $C(0)=\Delta^2(t)$, while $C(r)$ should vanish at large $r$. The functional form and the spatial extent of $C(r)$ contain important geometric information about the moving front~\cite{meakin1998fractals}. 

\begin{figure}
 \includegraphics[width=\linewidth]{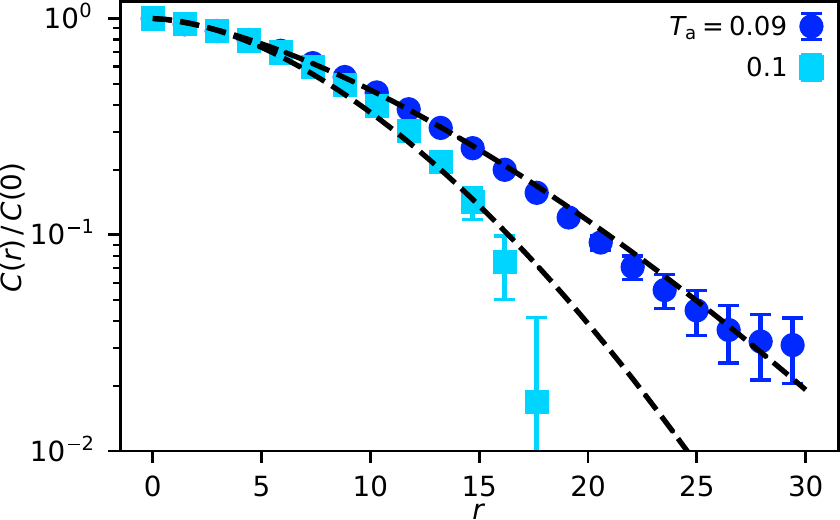}
 \caption{{\bf Spatial correlations of front position.}
   Normalized spatial correlation function $C(r)/C(0)$ from Eq.~(\ref{eq:Cr}) of the front position along the longitudinal $x$-direction for $T_i = 0.035$. Dashed lines correspond to a compressed exponential decay in Eq.~(\ref{eq:fit}) with a roughness exponent $H$ and a spatial correlation $\xi$ with $(H \approx 0.85, \xi \approx 10)$ for $T_a=0.1$ and $(H \approx 0.75, \xi \approx 12)$ for $T_a=0.09$.}
 \label{fig:fig7}
\end{figure}

In Fig.~\ref{fig:fig7} we show data for $C(r)/C(0)$ at two temperatures for which a steady state with a finite roughness can be reached. The spatial decay of $C(r)$ can be well-described by the functional form known for growing surfaces~\cite{meakin1998fractals,effect1993krim}:
\begin{equation}
C(r) = C(0) \exp \left[ - \left( \frac{r}{\xi} \right)^{2H} \right]  , \label{eq:fit}
\end{equation}
where $H$ is the roughness exponent and $\xi$ the longitudinal correlation length. This functional form quantifies both the spatial extent of the height-height correlations via the lengthscale $\xi$. The exponent $H$ quantifies how rough the interface is on distances shorter than $\xi$, because Eq.~(\ref{eq:fit}) leads to $\langle [ h(x) - h(x+r) ]^2 \rangle \sim r^{2H}$, so that $H=1$ for a very smooth interface. For the two temperatures shown we find $(H \approx 0.85, \xi \approx 10)$ for $T_a=0.1$ and $(H \approx 0.75, \xi \approx 12)$ for $T_a=0.09$.  

Overall, the mild temperature dependence of the interface characteristics suggests that as $T_a$ decreases the roughness $\Delta$ becomes smaller, the roughness exponent $H$ decreases, but the correlation length $\xi$ increases mildly suggesting that the front line becomes increasingly smooth (in the transverse directions) but with larger longitudinal correlations. Remarkably, the values that we extract for $\xi$ are in quantitative agreement with the dynamic lengthscales 
$\xi_4$ determined independently for this system in the bulk~\cite{thirty2022scalliet}, thus confirming our physical intuition that the corrugation of the moving interface retain the footprint of the spatially heterogeneous dynamics happening in the liquid.

Finally this direct connection between bulk relaxation and geometry of the front line allows us to also provide a microscopic explanation for the observed modest roughness of the propagating front in Fig.~\ref{fig:fig3}. In the time series of Fig.~\ref{fig:fig6} we notice that the front line resembles, at any fixed time, an elastic line with a succession of smooth maxima and minima. In this description minima correspond to regions where the front has moved less than the average in the immediate past. As a result, the glass region just above such region becomes surrounded on both sides by liquid regions where particle motion is now taking place much faster. This glass region is therefore dynamically facilitated on its left, right, and from below, and as a result its dynamics becomes much faster. This region is then very likely to relax in the near future. For the front line, this dynamic facilitation effect has therefore the tendency to reduce its roughness locally, as deep minima can in fact not survive over very long periods.  

\subsection{Why does the bulk diffusion constant control the velocity?}

We can now rationalize the observation that the front velocity $v_f$ is controlled by the diffusion constant $D$ of the supercooled liquid rather than by the structural relaxation time $\tau_\alpha$. While these quantities are proportional to one another in simple liquids as a result of the Stokes-Einstein relation, they decouple in supercooled liquids as a consequence of dynamic heterogeneity~\cite{ediger2000spatially}. The physical explanation, proposed long ago~\cite{ediger2000spatially,stillinger1994translation,tarjus1995} and refined in more recent microscopic approaches~\cite{LBerthier2005}, stems from the existence of a broad distribution of local relaxation times, $\pi(t)$, describing the spatial heterogeneity of the liquid structure. In this view, $\tau_\alpha$ receives more contributions from the largest times of the distribution, $\tau_\alpha \propto \langle t \rangle_\pi$, whereas the diffusion constant $D$ is dominated by the shortest times with $D \propto \langle 1/t \rangle_\pi$. For a narrow exponential distribution representative of a simple liquid, these two averages yield a similar timescale but they differ in a supercooled liquid with a broad underlying $\pi(t)$.  

In a very similar vein, our observations directly relate the local velocity of the front to the local relaxation time of the fluid as $v_f(x) \propto 1 / t(x)$, so that the spatially averaged velocity becomes $\langle 1/t \rangle_\pi$, just as the bulk diffusion constant. Therefore, we propose that, just as for $D$, the front velocity gives more weight to the short relaxation times of a broad underlying distribution. This argument rationalises the numerical finding in Fig.~\ref{fig:fig5}b that $v_f \sim D$ at low enough temperatures. Wolynes also proposed that the local relaxation rates control the local velocity~\cite{peter2009spatiotemporal}.

In this argument, the glass stability seems to play no role beyond the assumption that, on the timescale where the supercooled liquid relaxes, particles in the stable glass matrix are totally arrested and do not contribute to the front kinetics. However, we know from our simulations and from experiments that growth into a more stable glass is slower. We suggest that an increasing glass stability results in a denser, stiffer glass matrix into which the liquid must propagate, so that the growth should be slower, because more hindered. This effect can be quantified by constructing the following lengthscale 
\begin{equation}
\ell = \frac{v_f \sigma^2}{2 d  D},
\end{equation}
where $d$ is the space dimension. This length scale represents the averaged distance travelled by the front over the time $\sigma^2/(2dD)$ over which particles in the liquid diffuse a distance of one particle diameter $\sigma$. Our results show that $\ell$ is essentially a constant for sufficiently low $T_a$, and the lengthscale $\ell$ therefore usefully quantifies the (in)efficiency of the supercooled liquid at eroding particles in the stable glass. In other words, the ratio $\ell / \sigma$ appears as an efficient non-dimensional metric to quantify the kinetic stability of ultrastable glassy systems. Contrary to the stability ratio, for instance, $\ell$ does not depend on the temperature. 

For the three studied glasses, we find that $\ell / \sigma \approx 0.18$, 0.045, and 0.009 for increasing stability, thus varying by a factor of about 20, in agreement with the data shown in Fig.~\ref{fig:fig4}. This implies that for our most stable glasses, the front travels an average distance which about 100 times smaller than typical particle diffusion in the supercooled liquid, demonstrating how inefficient the supercooled liquid is at transforming the ultrastable glass. One could equivalently say that ultrastability precisely implies that the glass matrix is able to retain its structure despite the structural relaxation events taking place in the supercooled liquid nearby. This resistance to relaxation is captured by the lengthscale $\ell$.

Comparison with experiments on three-dimensional ultrastable molecular glasses indicate values of $\ell$ comparable to those found in our simulations. For TNB (tris-naphthylbenzene) and IMC (indomethacin), molecules with diameters of about 1 nm, values of $\ell$ are in the range $0.01 - 0.02$nm~\cite{sepulveda2012} and therefore $\ell /\sigma$ is also of order 1/100, as in our most stable computer glasses. 

\section{Discussion and proposed experiment}

\label{sec:conclusion}

Ultrastable glassy films not only represent a practically relevant new class of amorphous solids, they also offer novel research opportunities to resolve some of the outstanding questions surrounding the field of glass transition studies~\cite{berthier2016facets,ediger2017highly}. 

One of the most remarkable recent experimental findings is the observation that after a sudden heating, ultrastable glasses transform back into supercooled liquids very much as crystalline materials melt. Experiments are consistent with both heterogeneous and homogeneous nucleation kinetics, depending on experimental conditions. Together with thermodynamic measurements~\cite{ramos2011character,beasley2019vapor}, this set of experiments is in harmony with the possibility that a first-order phase transition underlies the liquid-glass transition~\cite{jack2016,guiselin2022glass}. 

\begin{figure}
 \includegraphics[width=\linewidth]{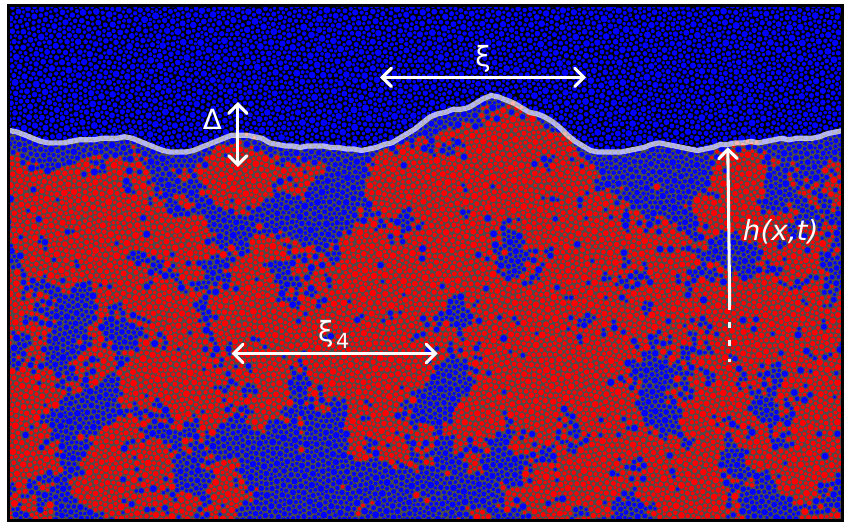}
 \caption{ { \bf Characteristic lengthscales.}
Snapshot illustrating the various lengthscales which characterise bulk liquid and front dynamics at $(Ti , Ta ) = (0.035, 0.085)$. The front is materialized by the white line at height $h(x,t)$, with a roughness characterized by $\Delta$ in Eq.~(\ref{eq:standard}). The longitudinal fluctuations of the interface decay over a lengthscale $\xi$ in Eq.~(\ref{eq:fit}) which directly reflects the extent of dynamic correlations in the bulk liquid $\xi_{4}$, thus offering a way to its direct experimental measurement. The size of the configuration is $160 \times 100$. Particles in the glass are shown in deep blue, while particles in the liquid are shown in red/blue according to the value of $C_B^i$ measured over a time window of duration $\tau_B$ preceding the snapshot.}
 \label{fig:fig8}
\end{figure}

Regarding the devitrification process, a major difference with crystal melting is that the obtained liquid is not a simple liquid relaxing homogeneously over microscopic timescales, but it is instead a very viscous supercooled liquid where structural relaxation is slow and spatially correlated. In a recent work, we have explored the consequences of these viscous dynamics when devitrification occurs in the bulk~\cite{herrero2022} and follows a peculiar type of nucleation and growth kinetics. 

Here, we focused on the heterogeneous process taking place in the presence of a macroscopic liquid-glass interface. This situation appears more natural from an experimental viewpoint as most ultrastable glassy systems are deposited films with a free interface which initiates the heterogeneous transformation. The combination of large-scale molecular dynamics simulations with the straightforward preparation of ultrastable initial configurations using the swap Monte Carlo algorithm allows us to rationalise most experimental findings while providing a microscopic interpretation of the observed phenomenology. In addition, our work offers detailed information about the geometry and space-time statistics of the propagating interface, which could be tested in future experiments.  

A major consequence of the unusual dynamics of the liquid phase is that the front velocity then depends very strongly on the annealing temperature $T_a$ at which the transformation takes place, and it can vary by many orders of magnitude, thus mirroring the dynamic slowing down of the bulk supercooled liquid. This result was well established experimentally, but our simulations provide essential insights into the microscopic processes underlying this macroscopic evolution. Our main result is the demonstration that the spatially heterogeneous dynamics of the bulk supercooled liquid is inherited by the propagating front which similarly moves in a spatially heterogeneous and temporally intermittent manner. A direct consequence is the prediction that the front propagation velocity is proportional to the diffusion constant of the fluid, $v_f \sim D$ and is thus decoupled from the structural relaxation time $\tau_\alpha$. 

We have also introduced two lengthscales characterising the spatial fluctuations of the propagating front, as summarised in Fig.~\ref{fig:fig8}. We have shown that the interface remains very smooth in the dynamic steady state, with a roughness in Eq.~(\ref{eq:standard}) which remains of the order of the molecular size and decreases at lower $T_a$. More interestingly perhaps, we showed that the longitudinal fluctuations of the interface display a larger correlation length scale, $\xi$, as a direct result of the underlying spatially correlated dynamics of the bulk liquids, as illustrated in Fig.~\ref{fig:fig8}. Our results and physical picture are consistent with the intuition that $\xi \approx \xi_4$, which suggests that the spatial fluctuations in the bulk relaxation dynamics leave their footprint in the shape of the propagating front.   

After decades of attempts, it remains notoriously difficult to directly measure $\xi_4$ in experiments on molecular liquids, as this would require resolving particle motion occurring at nanoscopic length scales over extremely large timescales. In fact, no technique can offer both spatial and temporal resolution. As a result, several efficient but somewhat indirect experimental estimates have instead been proposed to circumvent this challenge~\cite{tracht1998,reinsberg2001,berthier2005direct,dalle2007,zhang2018spatially}. 

We suggest that the geometry shown in Fig.~\ref{fig:fig8} offers an elegant solution to this problem, because the correlated particle motion occurring in the supercooled liquid leaves its signature on the geometry of the front line $h(x,t)$. In particular, we have shown that the longitudinal fluctuations of $h(x,t)$ are correlated over a lengthscale $\xi$ that is slaved to $\xi_4$. A crucial observation is that the front moves extremely slowly so that the detection of $\xi$ would require spatial resolution comparable to $\xi$ (but not necessarily $\sigma$), but temporal resolution of particle motion is no longer needed. Therefore we propose that an experiment designed to probe the spatial fluctuations of the interface separating the ultrastable glass from the transformed liquid would offer a direct experimental determination of the absolute value of $\xi$ over a range of annealing temperatures $T_a$, a result that has not been achieved so far. 

Although highly desirable, such an experiment is not necessarily straightforward to realise in practice. Since the front line separates two states of matter with distinct physical properties, it is potentially able to scatter light. For example, the density of the ultrastable glass and the supercooled liquid are different (typically of the order of 2\%~\cite{ediger2017highly}), which could be sufficient to scatter hard X-rays, which would provide excellent spatial resolution. The diffuse component of X-ray scattering~\cite{xray1988sinha} could then reveal the equal-time longitudinal fluctuations of $h(x,t)$ which are controlled by the roughness exponent $H$ and the correlation length $\xi$.  Another option is to prepare ultrastable glasses with elongated molecules since this can produce anisotropic glasses that are optically distinct from the isotropic supercooled liquid. Soft (resonant) X-rays~\cite{collins2012polarized} are sensitive to the front line of such systems~\cite{ferron2022characterization}, but spatial resolution may be somewhat limited due to their longer wavelength; this could be partially mitigated by using molecules containing an element with higher atomic number~\cite{freychet2021resonant}. We are optimistic that experts with scattering techniques can devise strategies to characterise the fluctuations of propagating fronts.

In the opening chapter of his famous book {\it The Little Prince}~\cite{prince}, Saint Exup\'ery shows a drawing made by the prince resembling an ordinary hat, but goes on to show that the contours of the `hat' actually reflect the shape of an elephant swallowed by a boa. Our results similarly show that the contour of the propagating front line contain a large amount of information about the correlated motion of the supercooled liquid taking place underneath. Another famous quote from the same book is that `what is essential is invisible to the eye'. This also metaphorically echoes our suggestion that an essential length scale of glass physics, that was so far invisible to the eye, may soon become accessible via a novel set of specifically devised  experiments. 

\acknowledgments

We thank R. Jack and D. DeLongchamp for useful discussions, and C. Scalliet for collaboration in the early stages of this project and careful reading of the manuscript. This work was publicly funded through ANR (the French National Research Agency) under the Investissements d'avenir programme with the reference ANR-16-IDEX-0006. It was also supported by a grant from the Simons Foundation (\#454933, LB), the U.S. National Science Foundation, CHE-2153944 (MDE), and by a Visiting Professorship from the Leverhulme Trust (VP1-2019-029, LB). 

\bibliography{cecilia.bib}

\begin{thebibliography}{71}%
\makeatletter
\providecommand \@ifxundefined [1]{%
 \@ifx{#1\undefined}
}%
\providecommand \@ifnum [1]{%
 \ifnum #1\expandafter \@firstoftwo
 \else \expandafter \@secondoftwo
 \fi
}%
\providecommand \@ifx [1]{%
 \ifx #1\expandafter \@firstoftwo
 \else \expandafter \@secondoftwo
 \fi
}%
\providecommand \natexlab [1]{#1}%
\providecommand \enquote  [1]{``#1''}%
\providecommand \bibnamefont  [1]{#1}%
\providecommand \bibfnamefont [1]{#1}%
\providecommand \citenamefont [1]{#1}%
\providecommand \href@noop [0]{\@secondoftwo}%
\providecommand \href [0]{\begingroup \@sanitize@url \@href}%
\providecommand \@href[1]{\@@startlink{#1}\@@href}%
\providecommand \@@href[1]{\endgroup#1\@@endlink}%
\providecommand \@sanitize@url [0]{\catcode `\\12\catcode `\$12\catcode
  `\&12\catcode `\#12\catcode `\^12\catcode `\_12\catcode `\%12\relax}%
\providecommand \@@startlink[1]{}%
\providecommand \@@endlink[0]{}%
\providecommand \url  [0]{\begingroup\@sanitize@url \@url }%
\providecommand \@url [1]{\endgroup\@href {#1}{\urlprefix }}%
\providecommand \urlprefix  [0]{URL }%
\providecommand \Eprint [0]{\href }%
\providecommand \doibase [0]{https://doi.org/}%
\providecommand \selectlanguage [0]{\@gobble}%
\providecommand \bibinfo  [0]{\@secondoftwo}%
\providecommand \bibfield  [0]{\@secondoftwo}%
\providecommand \translation [1]{[#1]}%
\providecommand \BibitemOpen [0]{}%
\providecommand \bibitemStop [0]{}%
\providecommand \bibitemNoStop [0]{.\EOS\space}%
\providecommand \EOS [0]{\spacefactor3000\relax}%
\providecommand \BibitemShut  [1]{\csname bibitem#1\endcsname}%
\let\auto@bib@innerbib\@empty
\bibitem [{\citenamefont {Berthier}\ and\ \citenamefont
  {Ediger}(2016)}]{berthier2016facets}%
  \BibitemOpen
  \bibfield  {author} {\bibinfo {author} {\bibfnamefont {L.}~\bibnamefont
  {Berthier}}\ and\ \bibinfo {author} {\bibfnamefont {M.~D.}\ \bibnamefont
  {Ediger}},\ }\bibfield  {title} {\bibinfo {title} {Facets of glass physics},\
  }\href {https://doi.org/10.1063/PT.3.3052} {\bibfield  {journal} {\bibinfo
  {journal} {Physics Today}\ }\textbf {\bibinfo {volume} {69}},\ \bibinfo
  {pages} {40} (\bibinfo {year} {2016})}\BibitemShut {NoStop}%
\bibitem [{\citenamefont {Debenedetti}\ and\ \citenamefont
  {Stillinger}(2001)}]{debenedetti2001supercooled}%
  \BibitemOpen
  \bibfield  {author} {\bibinfo {author} {\bibfnamefont {P.~G.}\ \bibnamefont
  {Debenedetti}}\ and\ \bibinfo {author} {\bibfnamefont {F.~H.}\ \bibnamefont
  {Stillinger}},\ }\bibfield  {title} {\bibinfo {title} {Supercooled liquids
  and the glass transition},\ }\href@noop {} {\bibfield  {journal} {\bibinfo
  {journal} {Nature}\ }\textbf {\bibinfo {volume} {410}},\ \bibinfo {pages}
  {259} (\bibinfo {year} {2001})}\BibitemShut {NoStop}%
\bibitem [{\citenamefont {Berthier}\ and\ \citenamefont
  {Biroli}(2011)}]{berthier2011theoretical}%
  \BibitemOpen
  \bibfield  {author} {\bibinfo {author} {\bibfnamefont {L.}~\bibnamefont
  {Berthier}}\ and\ \bibinfo {author} {\bibfnamefont {G.}~\bibnamefont
  {Biroli}},\ }\bibfield  {title} {\bibinfo {title} {Theoretical perspective on
  the glass transition and amorphous materials},\ }\href
  {https://doi.org/10.1103/RevModPhys.83.587} {\bibfield  {journal} {\bibinfo
  {journal} {Rev. Mod. Phys.}\ }\textbf {\bibinfo {volume} {83}},\ \bibinfo
  {pages} {587} (\bibinfo {year} {2011})}\BibitemShut {NoStop}%
\bibitem [{\citenamefont {Ediger}\ \emph {et~al.}(1996)\citenamefont {Ediger},
  \citenamefont {Angell},\ and\ \citenamefont {Nagel}}]{ediger1996supercooled}%
  \BibitemOpen
  \bibfield  {author} {\bibinfo {author} {\bibfnamefont {M.~D.}\ \bibnamefont
  {Ediger}}, \bibinfo {author} {\bibfnamefont {C.~A.}\ \bibnamefont {Angell}},\
  and\ \bibinfo {author} {\bibfnamefont {S.~R.}\ \bibnamefont {Nagel}},\
  }\bibfield  {title} {\bibinfo {title} {Supercooled liquids and glasses},\
  }\href {https://doi.org/10.1021/jp953538d} {\bibfield  {journal} {\bibinfo
  {journal} {The Journal of Physical Chemistry}\ }\textbf {\bibinfo {volume}
  {100}},\ \bibinfo {pages} {13200} (\bibinfo {year} {1996})}\BibitemShut
  {NoStop}%
\bibitem [{\citenamefont {Angell}\ \emph {et~al.}(2000)\citenamefont {Angell},
  \citenamefont {Ngai}, \citenamefont {McKenna}, \citenamefont {McMillan},\
  and\ \citenamefont {Martin}}]{angell2000relaxation}%
  \BibitemOpen
  \bibfield  {author} {\bibinfo {author} {\bibfnamefont {C.~A.}\ \bibnamefont
  {Angell}}, \bibinfo {author} {\bibfnamefont {K.~L.}\ \bibnamefont {Ngai}},
  \bibinfo {author} {\bibfnamefont {G.~B.}\ \bibnamefont {McKenna}}, \bibinfo
  {author} {\bibfnamefont {P.~F.}\ \bibnamefont {McMillan}},\ and\ \bibinfo
  {author} {\bibfnamefont {S.~W.}\ \bibnamefont {Martin}},\ }\bibfield  {title}
  {\bibinfo {title} {{Relaxation in glassforming liquids and amorphous
  solids}},\ }\href {https://doi.org/10.1063/1.1286035} {\bibfield  {journal}
  {\bibinfo  {journal} {Journal of Applied Physics}\ }\textbf {\bibinfo
  {volume} {88}},\ \bibinfo {pages} {3113} (\bibinfo {year}
  {2000})}\BibitemShut {NoStop}%
\bibitem [{\citenamefont {Ediger}(2000)}]{ediger2000spatially}%
  \BibitemOpen
  \bibfield  {author} {\bibinfo {author} {\bibfnamefont {M.~D.}\ \bibnamefont
  {Ediger}},\ }\bibfield  {title} {\bibinfo {title} {Spatially heterogeneous
  dynamics in supercooled liquids},\ }\href
  {https://doi.org/10.1146/annurev.physchem.51.1.99} {\bibfield  {journal}
  {\bibinfo  {journal} {Annual Review of Physical Chemistry}\ }\textbf
  {\bibinfo {volume} {51}},\ \bibinfo {pages} {99} (\bibinfo {year} {2000})},\
  \bibinfo {note} {pMID: 11031277}\BibitemShut {NoStop}%
\bibitem [{\citenamefont {Berthier}\ \emph {et~al.}(2011)\citenamefont
  {Berthier}, \citenamefont {Biroli}, \citenamefont {Bouchaud}, \citenamefont
  {Cipelletti},\ and\ \citenamefont {van Saarloos}}]{berthier2011dynamical}%
  \BibitemOpen
  \bibfield  {author} {\bibinfo {author} {\bibfnamefont {L.}~\bibnamefont
  {Berthier}}, \bibinfo {author} {\bibfnamefont {G.}~\bibnamefont {Biroli}},
  \bibinfo {author} {\bibfnamefont {J.-P.}\ \bibnamefont {Bouchaud}}, \bibinfo
  {author} {\bibfnamefont {L.}~\bibnamefont {Cipelletti}},\ and\ \bibinfo
  {author} {\bibfnamefont {W.}~\bibnamefont {van Saarloos}},\ }\href
  {https://doi.org/10.1093/acprof:oso/9780199691470.001.0001} {\emph {\bibinfo
  {title} {{Dynamical Heterogeneities in Glasses, Colloids, and Granular
  Media}}}}\ (\bibinfo  {publisher} {Oxford University Press},\ \bibinfo {year}
  {2011})\BibitemShut {NoStop}%
\bibitem [{\citenamefont {Swallen}\ \emph {et~al.}(2007)\citenamefont
  {Swallen}, \citenamefont {Kearns}, \citenamefont {Mapes}, \citenamefont
  {Kim}, \citenamefont {McMahon}, \citenamefont {Ediger}, \citenamefont {Wu},
  \citenamefont {Yu},\ and\ \citenamefont {Satija}}]{swallen2007}%
  \BibitemOpen
  \bibfield  {author} {\bibinfo {author} {\bibfnamefont {S.~F.}\ \bibnamefont
  {Swallen}}, \bibinfo {author} {\bibfnamefont {K.~L.}\ \bibnamefont {Kearns}},
  \bibinfo {author} {\bibfnamefont {M.~K.}\ \bibnamefont {Mapes}}, \bibinfo
  {author} {\bibfnamefont {Y.~S.}\ \bibnamefont {Kim}}, \bibinfo {author}
  {\bibfnamefont {R.~J.}\ \bibnamefont {McMahon}}, \bibinfo {author}
  {\bibfnamefont {M.~D.}\ \bibnamefont {Ediger}}, \bibinfo {author}
  {\bibfnamefont {T.}~\bibnamefont {Wu}}, \bibinfo {author} {\bibfnamefont
  {L.}~\bibnamefont {Yu}},\ and\ \bibinfo {author} {\bibfnamefont
  {S.}~\bibnamefont {Satija}},\ }\bibfield  {title} {\bibinfo {title} {Organic
  glasses with exceptional thermodynamic and kinetic stability},\ }\href
  {https://doi.org/10.1126/science.1135795} {\bibfield  {journal} {\bibinfo
  {journal} {Science}\ }\textbf {\bibinfo {volume} {315}},\ \bibinfo {pages}
  {353} (\bibinfo {year} {2007})}\BibitemShut {NoStop}%
\bibitem [{\citenamefont {Ediger}(2017)}]{ediger2017highly}%
  \BibitemOpen
  \bibfield  {author} {\bibinfo {author} {\bibfnamefont {M.~D.}\ \bibnamefont
  {Ediger}},\ }\bibfield  {title} {\bibinfo {title} {Perspective: Highly stable
  vapor-deposited glasses},\ }\href {https://doi.org/10.1063/1.5006265}
  {\bibfield  {journal} {\bibinfo  {journal} {The Journal of Chemical Physics}\
  }\textbf {\bibinfo {volume} {147}},\ \bibinfo {pages} {210901} (\bibinfo
  {year} {2017})}\BibitemShut {NoStop}%
\bibitem [{\citenamefont {Rodriguez-Tinoco}\ \emph {et~al.}(2022)\citenamefont
  {Rodriguez-Tinoco}, \citenamefont {Gonzalez-Silveira}, \citenamefont
  {Ramos},\ and\ \citenamefont {Rodriguez-Viejo}}]{rodriguez2022ultrastable}%
  \BibitemOpen
  \bibfield  {author} {\bibinfo {author} {\bibfnamefont {C.}~\bibnamefont
  {Rodriguez-Tinoco}}, \bibinfo {author} {\bibfnamefont {M.}~\bibnamefont
  {Gonzalez-Silveira}}, \bibinfo {author} {\bibfnamefont {M.~A.}\ \bibnamefont
  {Ramos}},\ and\ \bibinfo {author} {\bibfnamefont {J.}~\bibnamefont
  {Rodriguez-Viejo}},\ }\bibfield  {title} {\bibinfo {title} {Ultrastable
  glasses: new perspectives for an old problem},\ }\href
  {https://doi.org/10.1007/s40766-022-00029-y} {\bibfield  {journal} {\bibinfo
  {journal} {La Rivista del Nuovo Cimento}\ }\textbf {\bibinfo {volume} {45}},\
  \bibinfo {pages} {325} (\bibinfo {year} {2022})}\BibitemShut {NoStop}%
\bibitem [{\citenamefont {Liu}\ \emph {et~al.}(2015)\citenamefont {Liu},
  \citenamefont {Cheng}, \citenamefont {Salami-Ranjbaran}, \citenamefont {Gao},
  \citenamefont {Li}, \citenamefont {Tong}, \citenamefont {Lin}, \citenamefont
  {Zhang}, \citenamefont {Zhang}, \citenamefont {Klinge} \emph
  {et~al.}}]{liu2015}%
  \BibitemOpen
  \bibfield  {author} {\bibinfo {author} {\bibfnamefont {T.}~\bibnamefont
  {Liu}}, \bibinfo {author} {\bibfnamefont {K.}~\bibnamefont {Cheng}}, \bibinfo
  {author} {\bibfnamefont {E.}~\bibnamefont {Salami-Ranjbaran}}, \bibinfo
  {author} {\bibfnamefont {F.}~\bibnamefont {Gao}}, \bibinfo {author}
  {\bibfnamefont {C.}~\bibnamefont {Li}}, \bibinfo {author} {\bibfnamefont
  {X.}~\bibnamefont {Tong}}, \bibinfo {author} {\bibfnamefont {Y.-C.}\
  \bibnamefont {Lin}}, \bibinfo {author} {\bibfnamefont {Y.}~\bibnamefont
  {Zhang}}, \bibinfo {author} {\bibfnamefont {W.}~\bibnamefont {Zhang}},
  \bibinfo {author} {\bibfnamefont {L.}~\bibnamefont {Klinge}}, \emph
  {et~al.},\ }\bibfield  {title} {\bibinfo {title} {The effect of chemical
  structure on the stability of physical vapor deposited glasses of 1, 3,
  5-triarylbenzene},\ }\href {https://doi.org/10.1063/1.4928521} {\bibfield
  {journal} {\bibinfo  {journal} {The Journal of chemical physics}\ }\textbf
  {\bibinfo {volume} {143}},\ \bibinfo {pages} {084506} (\bibinfo {year}
  {2015})}\BibitemShut {NoStop}%
\bibitem [{\citenamefont {Tracht}\ \emph {et~al.}(1998)\citenamefont {Tracht},
  \citenamefont {Wilhelm}, \citenamefont {Heuer}, \citenamefont {Feng},
  \citenamefont {Schmidt-Rohr},\ and\ \citenamefont {Spiess}}]{tracht1998}%
  \BibitemOpen
  \bibfield  {author} {\bibinfo {author} {\bibfnamefont {U.}~\bibnamefont
  {Tracht}}, \bibinfo {author} {\bibfnamefont {M.}~\bibnamefont {Wilhelm}},
  \bibinfo {author} {\bibfnamefont {A.}~\bibnamefont {Heuer}}, \bibinfo
  {author} {\bibfnamefont {H.}~\bibnamefont {Feng}}, \bibinfo {author}
  {\bibfnamefont {K.}~\bibnamefont {Schmidt-Rohr}},\ and\ \bibinfo {author}
  {\bibfnamefont {H.~W.}\ \bibnamefont {Spiess}},\ }\bibfield  {title}
  {\bibinfo {title} {Length scale of dynamic heterogeneities at the glass
  transition determined by multidimensional nuclear magnetic resonance},\
  }\href {https://doi.org/10.1103/PhysRevLett.81.2727} {\bibfield  {journal}
  {\bibinfo  {journal} {Phys. Rev. Lett.}\ }\textbf {\bibinfo {volume} {81}},\
  \bibinfo {pages} {2727} (\bibinfo {year} {1998})}\BibitemShut {NoStop}%
\bibitem [{\citenamefont {Reinsberg}\ \emph {et~al.}(2001)\citenamefont
  {Reinsberg}, \citenamefont {Qiu}, \citenamefont {Wilhelm}, \citenamefont
  {Spiess},\ and\ \citenamefont {Ediger}}]{reinsberg2001}%
  \BibitemOpen
  \bibfield  {author} {\bibinfo {author} {\bibfnamefont {S.~A.}\ \bibnamefont
  {Reinsberg}}, \bibinfo {author} {\bibfnamefont {X.~H.}\ \bibnamefont {Qiu}},
  \bibinfo {author} {\bibfnamefont {M.}~\bibnamefont {Wilhelm}}, \bibinfo
  {author} {\bibfnamefont {H.~W.}\ \bibnamefont {Spiess}},\ and\ \bibinfo
  {author} {\bibfnamefont {M.~D.}\ \bibnamefont {Ediger}},\ }\bibfield  {title}
  {\bibinfo {title} {Length scale of dynamic heterogeneity in supercooled
  glycerol near $t_g$},\ }\href {https://doi.org/10.1063/1.1369160} {\bibfield
  {journal} {\bibinfo  {journal} {The Journal of Chemical Physics}\ }\textbf
  {\bibinfo {volume} {114}},\ \bibinfo {pages} {7299} (\bibinfo {year}
  {2001})}\BibitemShut {NoStop}%
\bibitem [{\citenamefont {Berthier}\ \emph {et~al.}(2005)\citenamefont
  {Berthier}, \citenamefont {Biroli}, \citenamefont {Bouchaud}, \citenamefont
  {Cipelletti}, \citenamefont {Masri}, \citenamefont {L'Hôte}, \citenamefont
  {Ladieu},\ and\ \citenamefont {Pierno}}]{berthier2005direct}%
  \BibitemOpen
  \bibfield  {author} {\bibinfo {author} {\bibfnamefont {L.}~\bibnamefont
  {Berthier}}, \bibinfo {author} {\bibfnamefont {G.}~\bibnamefont {Biroli}},
  \bibinfo {author} {\bibfnamefont {J.-P.}\ \bibnamefont {Bouchaud}}, \bibinfo
  {author} {\bibfnamefont {L.}~\bibnamefont {Cipelletti}}, \bibinfo {author}
  {\bibfnamefont {D.~E.}\ \bibnamefont {Masri}}, \bibinfo {author}
  {\bibfnamefont {D.}~\bibnamefont {L'Hôte}}, \bibinfo {author} {\bibfnamefont
  {F.}~\bibnamefont {Ladieu}},\ and\ \bibinfo {author} {\bibfnamefont
  {M.}~\bibnamefont {Pierno}},\ }\bibfield  {title} {\bibinfo {title} {Direct
  experimental evidence of a growing length scale accompanying the glass
  transition},\ }\href {https://doi.org/10.1126/science.1120714} {\bibfield
  {journal} {\bibinfo  {journal} {Science}\ }\textbf {\bibinfo {volume}
  {310}},\ \bibinfo {pages} {1797} (\bibinfo {year} {2005})}\BibitemShut
  {NoStop}%
\bibitem [{\citenamefont {Dalle-Ferrier}\ \emph {et~al.}(2007)\citenamefont
  {Dalle-Ferrier}, \citenamefont {Thibierge}, \citenamefont {Alba-Simionesco},
  \citenamefont {Berthier}, \citenamefont {Biroli}, \citenamefont {Bouchaud},
  \citenamefont {Ladieu}, \citenamefont {L'H\^ote},\ and\ \citenamefont
  {Tarjus}}]{dalle2007}%
  \BibitemOpen
  \bibfield  {author} {\bibinfo {author} {\bibfnamefont {C.}~\bibnamefont
  {Dalle-Ferrier}}, \bibinfo {author} {\bibfnamefont {C.}~\bibnamefont
  {Thibierge}}, \bibinfo {author} {\bibfnamefont {C.}~\bibnamefont
  {Alba-Simionesco}}, \bibinfo {author} {\bibfnamefont {L.}~\bibnamefont
  {Berthier}}, \bibinfo {author} {\bibfnamefont {G.}~\bibnamefont {Biroli}},
  \bibinfo {author} {\bibfnamefont {J.-P.}\ \bibnamefont {Bouchaud}}, \bibinfo
  {author} {\bibfnamefont {F.}~\bibnamefont {Ladieu}}, \bibinfo {author}
  {\bibfnamefont {D.}~\bibnamefont {L'H\^ote}},\ and\ \bibinfo {author}
  {\bibfnamefont {G.}~\bibnamefont {Tarjus}},\ }\bibfield  {title} {\bibinfo
  {title} {Spatial correlations in the dynamics of glassforming liquids:
  Experimental determination of their temperature dependence},\ }\href
  {https://doi.org/10.1103/PhysRevE.76.041510} {\bibfield  {journal} {\bibinfo
  {journal} {Phys. Rev. E}\ }\textbf {\bibinfo {volume} {76}},\ \bibinfo
  {pages} {041510} (\bibinfo {year} {2007})}\BibitemShut {NoStop}%
\bibitem [{\citenamefont {Zhang}\ \emph {et~al.}(2018)\citenamefont {Zhang},
  \citenamefont {Maldonis}, \citenamefont {Liu}, \citenamefont {Schroers},\
  and\ \citenamefont {Voyles}}]{zhang2018spatially}%
  \BibitemOpen
  \bibfield  {author} {\bibinfo {author} {\bibfnamefont {P.}~\bibnamefont
  {Zhang}}, \bibinfo {author} {\bibfnamefont {J.~J.}\ \bibnamefont {Maldonis}},
  \bibinfo {author} {\bibfnamefont {Z.}~\bibnamefont {Liu}}, \bibinfo {author}
  {\bibfnamefont {J.}~\bibnamefont {Schroers}},\ and\ \bibinfo {author}
  {\bibfnamefont {P.~M.}\ \bibnamefont {Voyles}},\ }\bibfield  {title}
  {\bibinfo {title} {Spatially heterogeneous dynamics in a metallic glass
  forming liquid imaged by electron correlation microscopy},\ }\href
  {https://doi.org/10.1038/s41467-018-03604-2} {\bibfield  {journal} {\bibinfo
  {journal} {Nature communications}\ }\textbf {\bibinfo {volume} {9}},\
  \bibinfo {pages} {1} (\bibinfo {year} {2018})}\BibitemShut {NoStop}%
\bibitem [{\citenamefont {Swallen}\ \emph {et~al.}(2009)\citenamefont
  {Swallen}, \citenamefont {Traynor}, \citenamefont {McMahon}, \citenamefont
  {Ediger},\ and\ \citenamefont {Mates}}]{swallen2009}%
  \BibitemOpen
  \bibfield  {author} {\bibinfo {author} {\bibfnamefont {S.~F.}\ \bibnamefont
  {Swallen}}, \bibinfo {author} {\bibfnamefont {K.}~\bibnamefont {Traynor}},
  \bibinfo {author} {\bibfnamefont {R.~J.}\ \bibnamefont {McMahon}}, \bibinfo
  {author} {\bibfnamefont {M.}~\bibnamefont {Ediger}},\ and\ \bibinfo {author}
  {\bibfnamefont {T.~E.}\ \bibnamefont {Mates}},\ }\bibfield  {title} {\bibinfo
  {title} {Stable glass transformation to supercooled liquid via
  surface-initiated growth front},\ }\href
  {https://doi.org/10.1103/PhysRevLett.102.065503} {\bibfield  {journal}
  {\bibinfo  {journal} {Physical review letters}\ }\textbf {\bibinfo {volume}
  {102}},\ \bibinfo {pages} {065503} (\bibinfo {year} {2009})}\BibitemShut
  {NoStop}%
\bibitem [{\citenamefont {Sep{\'u}lveda}\ \emph {et~al.}(2014)\citenamefont
  {Sep{\'u}lveda}, \citenamefont {Tylinski}, \citenamefont {Guiseppi-Elie},
  \citenamefont {Richert},\ and\ \citenamefont {Ediger}}]{sepulveda2014}%
  \BibitemOpen
  \bibfield  {author} {\bibinfo {author} {\bibfnamefont {A.}~\bibnamefont
  {Sep{\'u}lveda}}, \bibinfo {author} {\bibfnamefont {M.}~\bibnamefont
  {Tylinski}}, \bibinfo {author} {\bibfnamefont {A.}~\bibnamefont
  {Guiseppi-Elie}}, \bibinfo {author} {\bibfnamefont {R.}~\bibnamefont
  {Richert}},\ and\ \bibinfo {author} {\bibfnamefont {M.}~\bibnamefont
  {Ediger}},\ }\bibfield  {title} {\bibinfo {title} {Role of fragility in the
  formation of highly stable organic glasses},\ }\href
  {https://doi.org/10.1103/PhysRevLett.113.045901} {\bibfield  {journal}
  {\bibinfo  {journal} {Physical review letters}\ }\textbf {\bibinfo {volume}
  {113}},\ \bibinfo {pages} {045901} (\bibinfo {year} {2014})}\BibitemShut
  {NoStop}%
\bibitem [{\citenamefont {Rodr{\'\i}guez-Tinoco}\ \emph
  {et~al.}(2014)\citenamefont {Rodr{\'\i}guez-Tinoco}, \citenamefont
  {Gonzalez-Silveira}, \citenamefont {R{\`a}fols-Rib{\'e}}, \citenamefont
  {Lopeand{\'\i}a}, \citenamefont {Clavaguera-Mora},\ and\ \citenamefont
  {Rodr{\'\i}guez-Viejo}}]{rodriguez2014}%
  \BibitemOpen
  \bibfield  {author} {\bibinfo {author} {\bibfnamefont {C.}~\bibnamefont
  {Rodr{\'\i}guez-Tinoco}}, \bibinfo {author} {\bibfnamefont {M.}~\bibnamefont
  {Gonzalez-Silveira}}, \bibinfo {author} {\bibfnamefont {J.}~\bibnamefont
  {R{\`a}fols-Rib{\'e}}}, \bibinfo {author} {\bibfnamefont {A.~F.}\
  \bibnamefont {Lopeand{\'\i}a}}, \bibinfo {author} {\bibfnamefont {M.~T.}\
  \bibnamefont {Clavaguera-Mora}},\ and\ \bibinfo {author} {\bibfnamefont
  {J.}~\bibnamefont {Rodr{\'\i}guez-Viejo}},\ }\bibfield  {title} {\bibinfo
  {title} {Evaluation of growth front velocity in ultrastable glasses of
  indomethacin over a wide temperature interval},\ }\href
  {https://doi.org/10.1021/jp506782d} {\bibfield  {journal} {\bibinfo
  {journal} {The Journal of Physical Chemistry B}\ }\textbf {\bibinfo {volume}
  {118}},\ \bibinfo {pages} {10795} (\bibinfo {year} {2014})}\BibitemShut
  {NoStop}%
\bibitem [{\citenamefont {Hocky}\ \emph {et~al.}(2014)\citenamefont {Hocky},
  \citenamefont {Berthier},\ and\ \citenamefont {Reichman}}]{hocky2014}%
  \BibitemOpen
  \bibfield  {author} {\bibinfo {author} {\bibfnamefont {G.~M.}\ \bibnamefont
  {Hocky}}, \bibinfo {author} {\bibfnamefont {L.}~\bibnamefont {Berthier}},\
  and\ \bibinfo {author} {\bibfnamefont {D.~R.}\ \bibnamefont {Reichman}},\
  }\bibfield  {title} {\bibinfo {title} {Equilibrium ultrastable glasses
  produced by random pinning},\ }\href {https://doi.org/10.1063/1.4903200}
  {\bibfield  {journal} {\bibinfo  {journal} {The Journal of chemical physics}\
  }\textbf {\bibinfo {volume} {141}},\ \bibinfo {pages} {224503} (\bibinfo
  {year} {2014})}\BibitemShut {NoStop}%
\bibitem [{\citenamefont {Rodriguez-Tinoco}\ \emph {et~al.}(2019)\citenamefont
  {Rodriguez-Tinoco}, \citenamefont {Gonzalez-Silveira}, \citenamefont
  {Rafols-Ribe}, \citenamefont {Vila-Costa}, \citenamefont {Martinez-Garcia},\
  and\ \citenamefont {Rodriguez-Viejo}}]{rodriguez2019}%
  \BibitemOpen
  \bibfield  {author} {\bibinfo {author} {\bibfnamefont {C.}~\bibnamefont
  {Rodriguez-Tinoco}}, \bibinfo {author} {\bibfnamefont {M.}~\bibnamefont
  {Gonzalez-Silveira}}, \bibinfo {author} {\bibfnamefont {J.}~\bibnamefont
  {Rafols-Ribe}}, \bibinfo {author} {\bibfnamefont {A.}~\bibnamefont
  {Vila-Costa}}, \bibinfo {author} {\bibfnamefont {J.~C.}\ \bibnamefont
  {Martinez-Garcia}},\ and\ \bibinfo {author} {\bibfnamefont {J.}~\bibnamefont
  {Rodriguez-Viejo}},\ }\bibfield  {title} {\bibinfo {title} {Surface-bulk
  interplay in vapor-deposited glasses: Crossover length and the origin of
  front transformation},\ }\href
  {https://doi.org/10.1103/PhysRevLett.123.155501} {\bibfield  {journal}
  {\bibinfo  {journal} {Physical Review Letters}\ }\textbf {\bibinfo {volume}
  {123}},\ \bibinfo {pages} {155501} (\bibinfo {year} {2019})}\BibitemShut
  {NoStop}%
\bibitem [{\citenamefont {Flenner}\ \emph {et~al.}(2019)\citenamefont
  {Flenner}, \citenamefont {Berthier}, \citenamefont {Charbonneau},\ and\
  \citenamefont {Fullerton}}]{flenner2019}%
  \BibitemOpen
  \bibfield  {author} {\bibinfo {author} {\bibfnamefont {E.}~\bibnamefont
  {Flenner}}, \bibinfo {author} {\bibfnamefont {L.}~\bibnamefont {Berthier}},
  \bibinfo {author} {\bibfnamefont {P.}~\bibnamefont {Charbonneau}},\ and\
  \bibinfo {author} {\bibfnamefont {C.~J.}\ \bibnamefont {Fullerton}},\
  }\bibfield  {title} {\bibinfo {title} {Front-mediated melting of isotropic
  ultrastable glasses},\ }\href
  {https://doi.org/10.1103/PhysRevLett.123.175501} {\bibfield  {journal}
  {\bibinfo  {journal} {Physical review letters}\ }\textbf {\bibinfo {volume}
  {123}},\ \bibinfo {pages} {175501} (\bibinfo {year} {2019})}\BibitemShut
  {NoStop}%
\bibitem [{\citenamefont {Bhattacharya}\ and\ \citenamefont
  {Sadtchenko}(2014)}]{vlad2014}%
  \BibitemOpen
  \bibfield  {author} {\bibinfo {author} {\bibfnamefont {D.}~\bibnamefont
  {Bhattacharya}}\ and\ \bibinfo {author} {\bibfnamefont {V.}~\bibnamefont
  {Sadtchenko}},\ }\bibfield  {title} {\bibinfo {title} {Enthalpy and high
  temperature relaxation kinetics of stable vapor-deposited glasses of
  toluene},\ }\href {https://doi.org/10.1063/1.4893716} {\bibfield  {journal}
  {\bibinfo  {journal} {The Journal of Chemical Physics}\ }\textbf {\bibinfo
  {volume} {141}},\ \bibinfo {pages} {094502} (\bibinfo {year}
  {2014})}\BibitemShut {NoStop}%
\bibitem [{\citenamefont {Kearns}\ \emph {et~al.}(2010)\citenamefont {Kearns},
  \citenamefont {Ediger}, \citenamefont {Huth},\ and\ \citenamefont
  {Schick}}]{kearns2010}%
  \BibitemOpen
  \bibfield  {author} {\bibinfo {author} {\bibfnamefont {K.~L.}\ \bibnamefont
  {Kearns}}, \bibinfo {author} {\bibfnamefont {M.}~\bibnamefont {Ediger}},
  \bibinfo {author} {\bibfnamefont {H.}~\bibnamefont {Huth}},\ and\ \bibinfo
  {author} {\bibfnamefont {C.}~\bibnamefont {Schick}},\ }\bibfield  {title}
  {\bibinfo {title} {One micrometer length scale controls kinetic stability of
  low-energy glasses},\ }\href {https://doi.org/10.1021/jz9002179} {\bibfield
  {journal} {\bibinfo  {journal} {The Journal of Physical Chemistry Letters}\
  }\textbf {\bibinfo {volume} {1}},\ \bibinfo {pages} {388} (\bibinfo {year}
  {2010})}\BibitemShut {NoStop}%
\bibitem [{\citenamefont {Vila-Costa}\ \emph {et~al.}(2020)\citenamefont
  {Vila-Costa}, \citenamefont {R\`afols-Rib\'e}, \citenamefont
  {Gonz\'alez-Silveira}, \citenamefont {Lopeandia}, \citenamefont
  {Abad-Mu\~noz},\ and\ \citenamefont
  {Rodr\'{\i}guez-Viejo}}]{vila-costa2020nucleation}%
  \BibitemOpen
  \bibfield  {author} {\bibinfo {author} {\bibfnamefont {A.}~\bibnamefont
  {Vila-Costa}}, \bibinfo {author} {\bibfnamefont {J.}~\bibnamefont
  {R\`afols-Rib\'e}}, \bibinfo {author} {\bibfnamefont {M.}~\bibnamefont
  {Gonz\'alez-Silveira}}, \bibinfo {author} {\bibfnamefont {A.~F.}\
  \bibnamefont {Lopeandia}}, \bibinfo {author} {\bibfnamefont {L.}~\bibnamefont
  {Abad-Mu\~noz}},\ and\ \bibinfo {author} {\bibfnamefont {J.}~\bibnamefont
  {Rodr\'{\i}guez-Viejo}},\ }\bibfield  {title} {\bibinfo {title} {Nucleation
  and growth of the supercooled liquid phase control glass transition in bulk
  ultrastable glasses},\ }\href
  {https://doi.org/10.1103/PhysRevLett.124.076002} {\bibfield  {journal}
  {\bibinfo  {journal} {Phys. Rev. Lett.}\ }\textbf {\bibinfo {volume} {124}},\
  \bibinfo {pages} {076002} (\bibinfo {year} {2020})}\BibitemShut {NoStop}%
\bibitem [{\citenamefont {Fullerton}\ and\ \citenamefont
  {Berthier}(2017)}]{fullerton2017density}%
  \BibitemOpen
  \bibfield  {author} {\bibinfo {author} {\bibfnamefont {C.~J.}\ \bibnamefont
  {Fullerton}}\ and\ \bibinfo {author} {\bibfnamefont {L.}~\bibnamefont
  {Berthier}},\ }\bibfield  {title} {\bibinfo {title} {Density controls the
  kinetic stability of ultrastable glasses},\ }\href
  {https://doi.org/10.1209/0295-5075/119/36003} {\bibfield  {journal} {\bibinfo
   {journal} {Europhysics Letters}\ }\textbf {\bibinfo {volume} {119}},\
  \bibinfo {pages} {36003} (\bibinfo {year} {2017})}\BibitemShut {NoStop}%
\bibitem [{\citenamefont {Herrero}\ \emph {et~al.}(2023)\citenamefont
  {Herrero}, \citenamefont {Scalliet}, \citenamefont {Ediger},\ and\
  \citenamefont {Berthier}}]{herrero2022}%
  \BibitemOpen
  \bibfield  {author} {\bibinfo {author} {\bibfnamefont {C.}~\bibnamefont
  {Herrero}}, \bibinfo {author} {\bibfnamefont {C.}~\bibnamefont {Scalliet}},
  \bibinfo {author} {\bibfnamefont {M.~D.}\ \bibnamefont {Ediger}},\ and\
  \bibinfo {author} {\bibfnamefont {L.}~\bibnamefont {Berthier}},\ }\bibfield
  {title} {\bibinfo {title} {Two-step devitrification of ultrastable glasses},\
  }\href {https://doi.org/10.1073/pnas.2220824120} {\bibfield  {journal}
  {\bibinfo  {journal} {Proceedings of the National Academy of Sciences}\
  }\textbf {\bibinfo {volume} {120}},\ \bibinfo {pages} {e2220824120} (\bibinfo
  {year} {2023})}\BibitemShut {NoStop}%
\bibitem [{\citenamefont {Ruiz-Ruiz}\ \emph {et~al.}(2023)\citenamefont
  {Ruiz-Ruiz}, \citenamefont {Vila-Costa}, \citenamefont {Bar}, \citenamefont
  {Rodr{\'\i}guez-Tinoco}, \citenamefont {Gonzalez-Silveira}, \citenamefont
  {Plaza}, \citenamefont {Alcal{\'a}}, \citenamefont {Fraxedas},\ and\
  \citenamefont {Rodriguez-Viejo}}]{ruiz2023}%
  \BibitemOpen
  \bibfield  {author} {\bibinfo {author} {\bibfnamefont {M.}~\bibnamefont
  {Ruiz-Ruiz}}, \bibinfo {author} {\bibfnamefont {A.}~\bibnamefont
  {Vila-Costa}}, \bibinfo {author} {\bibfnamefont {T.}~\bibnamefont {Bar}},
  \bibinfo {author} {\bibfnamefont {C.}~\bibnamefont {Rodr{\'\i}guez-Tinoco}},
  \bibinfo {author} {\bibfnamefont {M.}~\bibnamefont {Gonzalez-Silveira}},
  \bibinfo {author} {\bibfnamefont {J.~A.}\ \bibnamefont {Plaza}}, \bibinfo
  {author} {\bibfnamefont {J.}~\bibnamefont {Alcal{\'a}}}, \bibinfo {author}
  {\bibfnamefont {J.}~\bibnamefont {Fraxedas}},\ and\ \bibinfo {author}
  {\bibfnamefont {J.}~\bibnamefont {Rodriguez-Viejo}},\ }\bibfield  {title}
  {\bibinfo {title} {Real-time microscopic view of the relaxation of a glass},\
  }\href {https://arxiv.org/abs/2301.08458} {\bibfield  {journal} {\bibinfo
  {journal} {arXiv preprint arXiv:2301.08458}\ } (\bibinfo {year}
  {2023})}\BibitemShut {NoStop}%
\bibitem [{\citenamefont {Vila-Costa}\ \emph {et~al.}(2023)\citenamefont
  {Vila-Costa}, \citenamefont {Gonzalez-Silveira}, \citenamefont
  {Rodr{\'\i}guez-Tinoco}, \citenamefont {Rodr{\'\i}guez-L{\'o}pez},\ and\
  \citenamefont {Rodriguez-Viejo}}]{vila2023}%
  \BibitemOpen
  \bibfield  {author} {\bibinfo {author} {\bibfnamefont {A.}~\bibnamefont
  {Vila-Costa}}, \bibinfo {author} {\bibfnamefont {M.}~\bibnamefont
  {Gonzalez-Silveira}}, \bibinfo {author} {\bibfnamefont {C.}~\bibnamefont
  {Rodr{\'\i}guez-Tinoco}}, \bibinfo {author} {\bibfnamefont {M.}~\bibnamefont
  {Rodr{\'\i}guez-L{\'o}pez}},\ and\ \bibinfo {author} {\bibfnamefont
  {J.}~\bibnamefont {Rodriguez-Viejo}},\ }\bibfield  {title} {\bibinfo {title}
  {Emergence of equilibrated liquid regions within the glass},\ }\href
  {https://doi.org/10.1038/s41567-022-01791-w} {\bibfield  {journal} {\bibinfo
  {journal} {Nature Physics}\ }\textbf {\bibinfo {volume} {19}},\ \bibinfo
  {pages} {114} (\bibinfo {year} {2023})}\BibitemShut {NoStop}%
\bibitem [{\citenamefont {Sepúlveda}\ \emph {et~al.}(2012)\citenamefont
  {Sepúlveda}, \citenamefont {Swallen}, \citenamefont {Kopff}, \citenamefont
  {McMahon},\ and\ \citenamefont {Ediger}}]{sepulveda2012}%
  \BibitemOpen
  \bibfield  {author} {\bibinfo {author} {\bibfnamefont {A.}~\bibnamefont
  {Sepúlveda}}, \bibinfo {author} {\bibfnamefont {S.~F.}\ \bibnamefont
  {Swallen}}, \bibinfo {author} {\bibfnamefont {L.~A.}\ \bibnamefont {Kopff}},
  \bibinfo {author} {\bibfnamefont {R.~J.}\ \bibnamefont {McMahon}},\ and\
  \bibinfo {author} {\bibfnamefont {M.~D.}\ \bibnamefont {Ediger}},\ }\bibfield
   {title} {\bibinfo {title} {Stable glasses of indomethacin and
  $\alpha$,$\alpha$,$\beta$-tris-naphthylbenzene transform into ordinary
  supercooled liquids},\ }\href {https://doi.org/10.1063/1.4768168} {\bibfield
  {journal} {\bibinfo  {journal} {The Journal of Chemical Physics}\ }\textbf
  {\bibinfo {volume} {137}},\ \bibinfo {pages} {204508} (\bibinfo {year}
  {2012})}\BibitemShut {NoStop}%
\bibitem [{\citenamefont {Rodríguez-Tinoco}\ \emph {et~al.}(2015)\citenamefont
  {Rodríguez-Tinoco}, \citenamefont {Gonzalez-Silveira}, \citenamefont
  {Ràfols-Ribé}, \citenamefont {Lopeandía},\ and\ \citenamefont
  {Rodríguez-Viejo}}]{rodriguez2015}%
  \BibitemOpen
  \bibfield  {author} {\bibinfo {author} {\bibfnamefont {C.}~\bibnamefont
  {Rodríguez-Tinoco}}, \bibinfo {author} {\bibfnamefont {M.}~\bibnamefont
  {Gonzalez-Silveira}}, \bibinfo {author} {\bibfnamefont {J.}~\bibnamefont
  {Ràfols-Ribé}}, \bibinfo {author} {\bibfnamefont {A.~F.}\ \bibnamefont
  {Lopeandía}},\ and\ \bibinfo {author} {\bibfnamefont {J.}~\bibnamefont
  {Rodríguez-Viejo}},\ }\bibfield  {title} {\bibinfo {title} {Transformation
  kinetics of vapor-deposited thin film organic glasses: the role of stability
  and molecular packing anisotropy},\ }\href
  {https://doi.org/10.1039/C5CP04692K} {\bibfield  {journal} {\bibinfo
  {journal} {Phys. Chem. Chem. Phys.}\ }\textbf {\bibinfo {volume} {17}},\
  \bibinfo {pages} {31195} (\bibinfo {year} {2015})}\BibitemShut {NoStop}%
\bibitem [{\citenamefont {Tylinski}\ \emph {et~al.}(2015)\citenamefont
  {Tylinski}, \citenamefont {Sep{\'u}lveda}, \citenamefont {Walters},
  \citenamefont {Chua}, \citenamefont {Schick},\ and\ \citenamefont
  {Ediger}}]{tylinski2015}%
  \BibitemOpen
  \bibfield  {author} {\bibinfo {author} {\bibfnamefont {M.}~\bibnamefont
  {Tylinski}}, \bibinfo {author} {\bibfnamefont {A.}~\bibnamefont
  {Sep{\'u}lveda}}, \bibinfo {author} {\bibfnamefont {D.~M.}\ \bibnamefont
  {Walters}}, \bibinfo {author} {\bibfnamefont {Y.}~\bibnamefont {Chua}},
  \bibinfo {author} {\bibfnamefont {C.}~\bibnamefont {Schick}},\ and\ \bibinfo
  {author} {\bibfnamefont {M.}~\bibnamefont {Ediger}},\ }\bibfield  {title}
  {\bibinfo {title} {Vapor-deposited glasses of methyl-m-toluate: How uniform
  is stable glass transformation?},\ }\href {https://doi.org/10.1063/1.4938420}
  {\bibfield  {journal} {\bibinfo  {journal} {The Journal of Chemical Physics}\
  }\textbf {\bibinfo {volume} {143}},\ \bibinfo {pages} {244509} (\bibinfo
  {year} {2015})}\BibitemShut {NoStop}%
\bibitem [{\citenamefont {Walters}\ \emph {et~al.}(2015)\citenamefont
  {Walters}, \citenamefont {Richert},\ and\ \citenamefont
  {Ediger}}]{walters2015}%
  \BibitemOpen
  \bibfield  {author} {\bibinfo {author} {\bibfnamefont {D.~M.}\ \bibnamefont
  {Walters}}, \bibinfo {author} {\bibfnamefont {R.}~\bibnamefont {Richert}},\
  and\ \bibinfo {author} {\bibfnamefont {M.~D.}\ \bibnamefont {Ediger}},\
  }\bibfield  {title} {\bibinfo {title} {Thermal stability of vapor-deposited
  stable glasses of an organic semiconductor},\ }\href
  {https://doi.org/10.1063/1.4916649} {\bibfield  {journal} {\bibinfo
  {journal} {The Journal of Chemical Physics}\ }\textbf {\bibinfo {volume}
  {142}},\ \bibinfo {pages} {134504} (\bibinfo {year} {2015})}\BibitemShut
  {NoStop}%
\bibitem [{\citenamefont {R{\`a}fols-Rib{\'e}}\ \emph
  {et~al.}(2017)\citenamefont {R{\`a}fols-Rib{\'e}}, \citenamefont
  {Gonzalez-Silveira}, \citenamefont {Rodr{\'\i}guez-Tinoco},\ and\
  \citenamefont {Rodr{\'\i}guez-Viejo}}]{rafols2017}%
  \BibitemOpen
  \bibfield  {author} {\bibinfo {author} {\bibfnamefont {J.}~\bibnamefont
  {R{\`a}fols-Rib{\'e}}}, \bibinfo {author} {\bibfnamefont {M.}~\bibnamefont
  {Gonzalez-Silveira}}, \bibinfo {author} {\bibfnamefont {C.}~\bibnamefont
  {Rodr{\'\i}guez-Tinoco}},\ and\ \bibinfo {author} {\bibfnamefont
  {J.}~\bibnamefont {Rodr{\'\i}guez-Viejo}},\ }\bibfield  {title} {\bibinfo
  {title} {The role of thermodynamic stability in the characteristics of the
  devitrification front of vapour-deposited glasses of toluene},\ }\href
  {https://doi.org/10.1039/C7CP00741H} {\bibfield  {journal} {\bibinfo
  {journal} {Physical Chemistry Chemical Physics}\ }\textbf {\bibinfo {volume}
  {19}},\ \bibinfo {pages} {11089} (\bibinfo {year} {2017})}\BibitemShut
  {NoStop}%
\bibitem [{\citenamefont {Dalal}\ and\ \citenamefont
  {Ediger}(2015)}]{dalal2015}%
  \BibitemOpen
  \bibfield  {author} {\bibinfo {author} {\bibfnamefont {S.~S.}\ \bibnamefont
  {Dalal}}\ and\ \bibinfo {author} {\bibfnamefont {M.~D.}\ \bibnamefont
  {Ediger}},\ }\bibfield  {title} {\bibinfo {title} {Influence of substrate
  temperature on the transformation front velocities that determine thermal
  stability of vapor-deposited glasses},\ }\href
  {https://doi.org/10.1021/jp512905a} {\bibfield  {journal} {\bibinfo
  {journal} {The Journal of Physical Chemistry B}\ }\textbf {\bibinfo {volume}
  {119}},\ \bibinfo {pages} {3875} (\bibinfo {year} {2015})}\BibitemShut
  {NoStop}%
\bibitem [{\citenamefont {Stillinger}\ and\ \citenamefont
  {Hodgdon}(1994)}]{stillinger1994translation}%
  \BibitemOpen
  \bibfield  {author} {\bibinfo {author} {\bibfnamefont {F.~H.}\ \bibnamefont
  {Stillinger}}\ and\ \bibinfo {author} {\bibfnamefont {J.~A.}\ \bibnamefont
  {Hodgdon}},\ }\bibfield  {title} {\bibinfo {title} {Translation-rotation
  paradox for diffusion in fragile glass-forming liquids},\ }\href
  {https://doi.org/10.1103/PhysRevE.50.2064} {\bibfield  {journal} {\bibinfo
  {journal} {Phys. Rev. E}\ }\textbf {\bibinfo {volume} {50}},\ \bibinfo
  {pages} {2064} (\bibinfo {year} {1994})}\BibitemShut {NoStop}%
\bibitem [{\citenamefont {Tarjus}\ and\ \citenamefont
  {Kivelson}(1995)}]{tarjus1995}%
  \BibitemOpen
  \bibfield  {author} {\bibinfo {author} {\bibfnamefont {G.}~\bibnamefont
  {Tarjus}}\ and\ \bibinfo {author} {\bibfnamefont {D.}~\bibnamefont
  {Kivelson}},\ }\bibfield  {title} {\bibinfo {title} {Breakdown of the
  stokes--einstein relation in supercooled liquids},\ }\href
  {https://doi.org/10.1063/1.470495} {\bibfield  {journal} {\bibinfo  {journal}
  {The Journal of chemical physics}\ }\textbf {\bibinfo {volume} {103}},\
  \bibinfo {pages} {3071} (\bibinfo {year} {1995})}\BibitemShut {NoStop}%
\bibitem [{\citenamefont {Whitaker}\ \emph {et~al.}(2012)\citenamefont
  {Whitaker}, \citenamefont {Ahrenberg}, \citenamefont {Schick},\ and\
  \citenamefont {Ediger}}]{whitaker2012}%
  \BibitemOpen
  \bibfield  {author} {\bibinfo {author} {\bibfnamefont {K.~R.}\ \bibnamefont
  {Whitaker}}, \bibinfo {author} {\bibfnamefont {M.}~\bibnamefont {Ahrenberg}},
  \bibinfo {author} {\bibfnamefont {C.}~\bibnamefont {Schick}},\ and\ \bibinfo
  {author} {\bibfnamefont {M.}~\bibnamefont {Ediger}},\ }\bibfield  {title}
  {\bibinfo {title} {Vapor-deposited $\alpha$, $\alpha$,
  $\beta$-tris-naphthylbenzene glasses with low heat capacity and high kinetic
  stability},\ }\href {https://doi.org/10.1063/1.4758807} {\bibfield  {journal}
  {\bibinfo  {journal} {The Journal of Chemical Physics}\ }\textbf {\bibinfo
  {volume} {137}},\ \bibinfo {pages} {154502} (\bibinfo {year}
  {2012})}\BibitemShut {NoStop}%
\bibitem [{\citenamefont {Kaur}\ \emph {et~al.}(2023)\citenamefont {Kaur},
  \citenamefont {Bhattacharya}, \citenamefont {Cubeta},\ and\ \citenamefont
  {Sadtchenko}}]{vlad2023}%
  \BibitemOpen
  \bibfield  {author} {\bibinfo {author} {\bibfnamefont {R.}~\bibnamefont
  {Kaur}}, \bibinfo {author} {\bibfnamefont {D.}~\bibnamefont {Bhattacharya}},
  \bibinfo {author} {\bibfnamefont {U.}~\bibnamefont {Cubeta}},\ and\ \bibinfo
  {author} {\bibfnamefont {V.}~\bibnamefont {Sadtchenko}},\ }\bibfield  {title}
  {\bibinfo {title} {Glass softening in the limit of high heating rates:
  heterogeneous devitrification kinetics on nano, meso, and micrometer scale},\
  }\href@noop {} {\bibfield  {journal} {\bibinfo  {journal} {in preparation}\ }
  (\bibinfo {year} {2023})}\BibitemShut {NoStop}%
\bibitem [{\citenamefont {Fredrickson}\ and\ \citenamefont
  {Andersen}(1984)}]{fredrickson1984kinetic}%
  \BibitemOpen
  \bibfield  {author} {\bibinfo {author} {\bibfnamefont {G.~H.}\ \bibnamefont
  {Fredrickson}}\ and\ \bibinfo {author} {\bibfnamefont {H.~C.}\ \bibnamefont
  {Andersen}},\ }\bibfield  {title} {\bibinfo {title} {Kinetic ising model of
  the glass transition},\ }\href {https://doi.org/10.1103/PhysRevLett.53.1244}
  {\bibfield  {journal} {\bibinfo  {journal} {Phys. Rev. Lett.}\ }\textbf
  {\bibinfo {volume} {53}},\ \bibinfo {pages} {1244} (\bibinfo {year}
  {1984})}\BibitemShut {NoStop}%
\bibitem [{\citenamefont {Chandler}\ and\ \citenamefont
  {Garrahan}(2010)}]{chandler2010dynamics}%
  \BibitemOpen
  \bibfield  {author} {\bibinfo {author} {\bibfnamefont {D.}~\bibnamefont
  {Chandler}}\ and\ \bibinfo {author} {\bibfnamefont {J.~P.}\ \bibnamefont
  {Garrahan}},\ }\bibfield  {title} {\bibinfo {title} {Dynamics on the way to
  forming glass: Bubbles in space-time},\ }\href
  {https://doi.org/10.1146/annurev.physchem.040808.090405} {\bibfield
  {journal} {\bibinfo  {journal} {Annual Review of Physical Chemistry}\
  }\textbf {\bibinfo {volume} {61}},\ \bibinfo {pages} {191} (\bibinfo {year}
  {2010})}\BibitemShut {NoStop}%
\bibitem [{\citenamefont {Sepúlveda}\ \emph {et~al.}(2013)\citenamefont
  {Sepúlveda}, \citenamefont {Swallen},\ and\ \citenamefont
  {Ediger}}]{sepulveda2014manipulating}%
  \BibitemOpen
  \bibfield  {author} {\bibinfo {author} {\bibfnamefont {A.}~\bibnamefont
  {Sepúlveda}}, \bibinfo {author} {\bibfnamefont {S.~F.}\ \bibnamefont
  {Swallen}},\ and\ \bibinfo {author} {\bibfnamefont {M.~D.}\ \bibnamefont
  {Ediger}},\ }\bibfield  {title} {\bibinfo {title} {Manipulating the
  properties of stable organic glasses using kinetic facilitation},\ }\href
  {https://doi.org/10.1063/1.4772594} {\bibfield  {journal} {\bibinfo
  {journal} {The Journal of Chemical Physics}\ }\textbf {\bibinfo {volume}
  {138}},\ \bibinfo {pages} {12A517} (\bibinfo {year} {2013})}\BibitemShut
  {NoStop}%
\bibitem [{\citenamefont {L{\'e}onard}\ and\ \citenamefont
  {Harrowell}(2010)}]{leonard2010}%
  \BibitemOpen
  \bibfield  {author} {\bibinfo {author} {\bibfnamefont {S.}~\bibnamefont
  {L{\'e}onard}}\ and\ \bibinfo {author} {\bibfnamefont {P.}~\bibnamefont
  {Harrowell}},\ }\bibfield  {title} {\bibinfo {title} {Macroscopic
  facilitation of glassy relaxation kinetics: Ultrastable glass films with
  frontlike thermal response},\ }\href {https://doi.org/10.1063/1.3511721}
  {\bibfield  {journal} {\bibinfo  {journal} {The Journal of chemical physics}\
  }\textbf {\bibinfo {volume} {133}},\ \bibinfo {pages} {244502} (\bibinfo
  {year} {2010})}\BibitemShut {NoStop}%
\bibitem [{\citenamefont {Guti{\'e}rrez}\ and\ \citenamefont
  {Garrahan}(2016)}]{gutierrez2016}%
  \BibitemOpen
  \bibfield  {author} {\bibinfo {author} {\bibfnamefont {R.}~\bibnamefont
  {Guti{\'e}rrez}}\ and\ \bibinfo {author} {\bibfnamefont {J.~P.}\ \bibnamefont
  {Garrahan}},\ }\bibfield  {title} {\bibinfo {title} {Front propagation versus
  bulk relaxation in the annealing dynamics of a kinetically constrained model
  of ultrastable glasses},\ }\href
  {https://doi.org/10.1088/1742-5468/2016/07/074005} {\bibfield  {journal}
  {\bibinfo  {journal} {Journal of Statistical Mechanics: Theory and
  Experiment}\ }\textbf {\bibinfo {volume} {2016}},\ \bibinfo {pages} {074005}
  (\bibinfo {year} {2016})}\BibitemShut {NoStop}%
\bibitem [{\citenamefont {Wolynes}(2009)}]{peter2009spatiotemporal}%
  \BibitemOpen
  \bibfield  {author} {\bibinfo {author} {\bibfnamefont {P.~G.}\ \bibnamefont
  {Wolynes}},\ }\bibfield  {title} {\bibinfo {title} {Spatiotemporal structures
  in aging and rejuvenating glasses},\ }\href
  {https://doi.org/10.1073/pnas.0812418106} {\bibfield  {journal} {\bibinfo
  {journal} {Proceedings of the National Academy of Sciences}\ }\textbf
  {\bibinfo {volume} {106}},\ \bibinfo {pages} {1353} (\bibinfo {year}
  {2009})}\BibitemShut {NoStop}%
\bibitem [{\citenamefont {Wisitsorasak}\ and\ \citenamefont
  {Wolynes}(2013)}]{wisitsorasak2013}%
  \BibitemOpen
  \bibfield  {author} {\bibinfo {author} {\bibfnamefont {A.}~\bibnamefont
  {Wisitsorasak}}\ and\ \bibinfo {author} {\bibfnamefont {P.~G.}\ \bibnamefont
  {Wolynes}},\ }\bibfield  {title} {\bibinfo {title} {Fluctuating mobility
  generation and transport in glasses},\ }\href
  {https://doi.org/10.1103/PhysRevE.88.022308} {\bibfield  {journal} {\bibinfo
  {journal} {Physical Review E}\ }\textbf {\bibinfo {volume} {88}},\ \bibinfo
  {pages} {022308} (\bibinfo {year} {2013})}\BibitemShut {NoStop}%
\bibitem [{\citenamefont {Berthier}\ and\ \citenamefont
  {Reichman}(2023)}]{berthier2023modern}%
  \BibitemOpen
  \bibfield  {author} {\bibinfo {author} {\bibfnamefont {L.}~\bibnamefont
  {Berthier}}\ and\ \bibinfo {author} {\bibfnamefont {D.~R.}\ \bibnamefont
  {Reichman}},\ }\bibfield  {title} {\bibinfo {title} {Modern computational
  studies of the glass transition},\ }\href
  {https://doi.org/10.1038/s42254-022-00548-x} {\bibfield  {journal} {\bibinfo
  {journal} {Nature Reviews Physics}\ ,\ \bibinfo {pages} {1}} (\bibinfo {year}
  {2023})}\BibitemShut {NoStop}%
\bibitem [{\citenamefont {Lyubimov}\ \emph {et~al.}(2013)\citenamefont
  {Lyubimov}, \citenamefont {Ediger},\ and\ \citenamefont
  {de~Pablo}}]{lyubimov2013}%
  \BibitemOpen
  \bibfield  {author} {\bibinfo {author} {\bibfnamefont {I.}~\bibnamefont
  {Lyubimov}}, \bibinfo {author} {\bibfnamefont {M.~D.}\ \bibnamefont
  {Ediger}},\ and\ \bibinfo {author} {\bibfnamefont {J.~J.}\ \bibnamefont
  {de~Pablo}},\ }\bibfield  {title} {\bibinfo {title} {Model vapor-deposited
  glasses: Growth front and composition effects},\ }\href
  {https://doi.org/10.1063/1.4823769} {\bibfield  {journal} {\bibinfo
  {journal} {The Journal of Chemical Physics}\ }\textbf {\bibinfo {volume}
  {139}},\ \bibinfo {pages} {144505} (\bibinfo {year} {2013})}\BibitemShut
  {NoStop}%
\bibitem [{\citenamefont {Berthier}\ \emph {et~al.}(2016)\citenamefont
  {Berthier}, \citenamefont {Coslovich}, \citenamefont {Ninarello},\ and\
  \citenamefont {Ozawa}}]{equilibrium2016ozawa}%
  \BibitemOpen
  \bibfield  {author} {\bibinfo {author} {\bibfnamefont {L.}~\bibnamefont
  {Berthier}}, \bibinfo {author} {\bibfnamefont {D.}~\bibnamefont {Coslovich}},
  \bibinfo {author} {\bibfnamefont {A.}~\bibnamefont {Ninarello}},\ and\
  \bibinfo {author} {\bibfnamefont {M.}~\bibnamefont {Ozawa}},\ }\bibfield
  {title} {\bibinfo {title} {Equilibrium sampling of hard spheres up to the
  jamming density and beyond},\ }\href
  {https://doi.org/10.1103/PhysRevLett.116.238002} {\bibfield  {journal}
  {\bibinfo  {journal} {Phys. Rev. Lett.}\ }\textbf {\bibinfo {volume} {116}},\
  \bibinfo {pages} {238002} (\bibinfo {year} {2016})}\BibitemShut {NoStop}%
\bibitem [{\citenamefont {Ninarello}\ \emph {et~al.}(2017)\citenamefont
  {Ninarello}, \citenamefont {Berthier},\ and\ \citenamefont
  {Coslovich}}]{ninarello2017}%
  \BibitemOpen
  \bibfield  {author} {\bibinfo {author} {\bibfnamefont {A.}~\bibnamefont
  {Ninarello}}, \bibinfo {author} {\bibfnamefont {L.}~\bibnamefont
  {Berthier}},\ and\ \bibinfo {author} {\bibfnamefont {D.}~\bibnamefont
  {Coslovich}},\ }\bibfield  {title} {\bibinfo {title} {Models and algorithms
  for the next generation of glass transition studies},\ }\href
  {https://doi.org/10.1103/PhysRevX.7.021039} {\bibfield  {journal} {\bibinfo
  {journal} {Physical Review X}\ }\textbf {\bibinfo {volume} {7}},\ \bibinfo
  {pages} {021039} (\bibinfo {year} {2017})}\BibitemShut {NoStop}%
\bibitem [{\citenamefont {Berthier}\ \emph
  {et~al.}(2019{\natexlab{a}})\citenamefont {Berthier}, \citenamefont
  {Flenner}, \citenamefont {Fullerton}, \citenamefont {Scalliet},\ and\
  \citenamefont {Singh}}]{berthier2019swap}%
  \BibitemOpen
  \bibfield  {author} {\bibinfo {author} {\bibfnamefont {L.}~\bibnamefont
  {Berthier}}, \bibinfo {author} {\bibfnamefont {E.}~\bibnamefont {Flenner}},
  \bibinfo {author} {\bibfnamefont {C.~J.}\ \bibnamefont {Fullerton}}, \bibinfo
  {author} {\bibfnamefont {C.}~\bibnamefont {Scalliet}},\ and\ \bibinfo
  {author} {\bibfnamefont {M.}~\bibnamefont {Singh}},\ }\bibfield  {title}
  {\bibinfo {title} {Efficient swap algorithms for molecular dynamics
  simulations of equilibrium supercooled liquids},\ }\href
  {https://doi.org/10.1088/1742-5468/ab1910} {\bibfield  {journal} {\bibinfo
  {journal} {Journal of Statistical Mechanics: Theory and Experiment}\ }\textbf
  {\bibinfo {volume} {2019}},\ \bibinfo {pages} {064004} (\bibinfo {year}
  {2019}{\natexlab{a}})}\BibitemShut {NoStop}%
\bibitem [{\citenamefont {Sinha}\ \emph {et~al.}(1988)\citenamefont {Sinha},
  \citenamefont {Sirota}, \citenamefont {Garoff},\ and\ \citenamefont
  {Stanley}}]{xray1988sinha}%
  \BibitemOpen
  \bibfield  {author} {\bibinfo {author} {\bibfnamefont {S.~K.}\ \bibnamefont
  {Sinha}}, \bibinfo {author} {\bibfnamefont {E.~B.}\ \bibnamefont {Sirota}},
  \bibinfo {author} {\bibfnamefont {S.}~\bibnamefont {Garoff}},\ and\ \bibinfo
  {author} {\bibfnamefont {H.~B.}\ \bibnamefont {Stanley}},\ }\bibfield
  {title} {\bibinfo {title} {X-ray and neutron scattering from rough
  surfaces},\ }\href {https://doi.org/10.1103/PhysRevB.38.2297} {\bibfield
  {journal} {\bibinfo  {journal} {Phys. Rev. B}\ }\textbf {\bibinfo {volume}
  {38}},\ \bibinfo {pages} {2297} (\bibinfo {year} {1988})}\BibitemShut
  {NoStop}%
\bibitem [{\citenamefont {Berthier}\ \emph
  {et~al.}(2019{\natexlab{b}})\citenamefont {Berthier}, \citenamefont
  {Charbonneau}, \citenamefont {Ninarello}, \citenamefont {Ozawa},\ and\
  \citenamefont {Yaida}}]{berthier2019zero}%
  \BibitemOpen
  \bibfield  {author} {\bibinfo {author} {\bibfnamefont {L.}~\bibnamefont
  {Berthier}}, \bibinfo {author} {\bibfnamefont {P.}~\bibnamefont
  {Charbonneau}}, \bibinfo {author} {\bibfnamefont {A.}~\bibnamefont
  {Ninarello}}, \bibinfo {author} {\bibfnamefont {M.}~\bibnamefont {Ozawa}},\
  and\ \bibinfo {author} {\bibfnamefont {S.}~\bibnamefont {Yaida}},\ }\bibfield
   {title} {\bibinfo {title} {Zero-temperature glass transition in two
  dimensions},\ }\href {https://doi.org/10.1038/s41467-019-09512-3} {\bibfield
  {journal} {\bibinfo  {journal} {Nature communications}\ }\textbf {\bibinfo
  {volume} {10}},\ \bibinfo {pages} {1508} (\bibinfo {year}
  {2019}{\natexlab{b}})}\BibitemShut {NoStop}%
\bibitem [{\citenamefont {Guiselin}\ \emph
  {et~al.}(2022{\natexlab{a}})\citenamefont {Guiselin}, \citenamefont
  {Scalliet},\ and\ \citenamefont {Berthier}}]{guiselin2022microscopic}%
  \BibitemOpen
  \bibfield  {author} {\bibinfo {author} {\bibfnamefont {B.}~\bibnamefont
  {Guiselin}}, \bibinfo {author} {\bibfnamefont {C.}~\bibnamefont {Scalliet}},\
  and\ \bibinfo {author} {\bibfnamefont {L.}~\bibnamefont {Berthier}},\
  }\bibfield  {title} {\bibinfo {title} {Microscopic origin of excess wings in
  relaxation spectra of supercooled liquids},\ }\href
  {https://doi.org/10.1038/s41567-022-01508-z} {\bibfield  {journal} {\bibinfo
  {journal} {Nature Physics}\ }\textbf {\bibinfo {volume} {18}},\ \bibinfo
  {pages} {468} (\bibinfo {year} {2022}{\natexlab{a}})}\BibitemShut {NoStop}%
\bibitem [{\citenamefont {Scalliet}\ \emph {et~al.}(2022)\citenamefont
  {Scalliet}, \citenamefont {Guiselin},\ and\ \citenamefont
  {Berthier}}]{thirty2022scalliet}%
  \BibitemOpen
  \bibfield  {author} {\bibinfo {author} {\bibfnamefont {C.}~\bibnamefont
  {Scalliet}}, \bibinfo {author} {\bibfnamefont {B.}~\bibnamefont {Guiselin}},\
  and\ \bibinfo {author} {\bibfnamefont {L.}~\bibnamefont {Berthier}},\
  }\bibfield  {title} {\bibinfo {title} {Thirty milliseconds in the life of a
  supercooled liquid},\ }\href {https://doi.org/10.1103/PhysRevX.12.041028}
  {\bibfield  {journal} {\bibinfo  {journal} {Phys. Rev. X}\ }\textbf {\bibinfo
  {volume} {12}},\ \bibinfo {pages} {041028} (\bibinfo {year}
  {2022})}\BibitemShut {NoStop}%
\bibitem [{\citenamefont {Barabási}\ and\ \citenamefont
  {Stanley}(1995)}]{barabasi1995fractal}%
  \BibitemOpen
  \bibfield  {author} {\bibinfo {author} {\bibfnamefont {A.-L.}\ \bibnamefont
  {Barabási}}\ and\ \bibinfo {author} {\bibfnamefont {H.~E.}\ \bibnamefont
  {Stanley}},\ }\href {https://doi.org/10.1017/CBO9780511599798} {\emph
  {\bibinfo {title} {Fractal Concepts in Surface Growth}}}\ (\bibinfo
  {publisher} {Cambridge University Press},\ \bibinfo {year}
  {1995})\BibitemShut {NoStop}%
\bibitem [{\citenamefont {Meakin}(1998)}]{meakin1998fractals}%
  \BibitemOpen
  \bibfield  {author} {\bibinfo {author} {\bibfnamefont {P.}~\bibnamefont
  {Meakin}},\ }\href@noop {} {\emph {\bibinfo {title} {Fractals, scaling and
  growth far from equilibrium}}},\ Vol.~\bibinfo {volume} {5}\ (\bibinfo
  {publisher} {Cambridge university press},\ \bibinfo {year}
  {1998})\BibitemShut {NoStop}%
\bibitem [{\citenamefont {Berthier}\ and\ \citenamefont
  {Ediger}(2020)}]{how2020berthier}%
  \BibitemOpen
  \bibfield  {author} {\bibinfo {author} {\bibfnamefont {L.}~\bibnamefont
  {Berthier}}\ and\ \bibinfo {author} {\bibfnamefont {M.~D.}\ \bibnamefont
  {Ediger}},\ }\bibfield  {title} {\bibinfo {title} {How to “measure” a
  structural relaxation time that is too long to be measured?},\ }\href
  {https://doi.org/10.1063/5.0015227} {\bibfield  {journal} {\bibinfo
  {journal} {The Journal of Chemical Physics}\ }\textbf {\bibinfo {volume}
  {153}},\ \bibinfo {pages} {044501} (\bibinfo {year} {2020})}\BibitemShut
  {NoStop}%
\bibitem [{\citenamefont {Illing}\ \emph {et~al.}(2017)\citenamefont {Illing},
  \citenamefont {Fritschi}, \citenamefont {Kaiser}, \citenamefont {Klix},
  \citenamefont {Maret},\ and\ \citenamefont {Keim}}]{illing2017}%
  \BibitemOpen
  \bibfield  {author} {\bibinfo {author} {\bibfnamefont {B.}~\bibnamefont
  {Illing}}, \bibinfo {author} {\bibfnamefont {S.}~\bibnamefont {Fritschi}},
  \bibinfo {author} {\bibfnamefont {H.}~\bibnamefont {Kaiser}}, \bibinfo
  {author} {\bibfnamefont {C.~L.}\ \bibnamefont {Klix}}, \bibinfo {author}
  {\bibfnamefont {G.}~\bibnamefont {Maret}},\ and\ \bibinfo {author}
  {\bibfnamefont {P.}~\bibnamefont {Keim}},\ }\bibfield  {title} {\bibinfo
  {title} {Mermin–wagner fluctuations in 2d amorphous solids},\ }\href
  {https://doi.org/10.1073/pnas.1612964114} {\bibfield  {journal} {\bibinfo
  {journal} {Proceedings of the National Academy of Sciences}\ }\textbf
  {\bibinfo {volume} {114}},\ \bibinfo {pages} {1856} (\bibinfo {year}
  {2017})}\BibitemShut {NoStop}%
\bibitem [{\citenamefont {Kob}\ \emph {et~al.}(1997)\citenamefont {Kob},
  \citenamefont {Donati}, \citenamefont {Plimpton}, \citenamefont {Poole},\
  and\ \citenamefont {Glotzer}}]{kob1997dynamical}%
  \BibitemOpen
  \bibfield  {author} {\bibinfo {author} {\bibfnamefont {W.}~\bibnamefont
  {Kob}}, \bibinfo {author} {\bibfnamefont {C.}~\bibnamefont {Donati}},
  \bibinfo {author} {\bibfnamefont {S.~J.}\ \bibnamefont {Plimpton}}, \bibinfo
  {author} {\bibfnamefont {P.~H.}\ \bibnamefont {Poole}},\ and\ \bibinfo
  {author} {\bibfnamefont {S.~C.}\ \bibnamefont {Glotzer}},\ }\bibfield
  {title} {\bibinfo {title} {Dynamical heterogeneities in a supercooled
  lennard-jones liquid},\ }\href {https://doi.org/10.1103/PhysRevLett.79.2827}
  {\bibfield  {journal} {\bibinfo  {journal} {Phys. Rev. Lett.}\ }\textbf
  {\bibinfo {volume} {79}},\ \bibinfo {pages} {2827} (\bibinfo {year}
  {1997})}\BibitemShut {NoStop}%
\bibitem [{\citenamefont {Chaudhuri}\ \emph {et~al.}(2007)\citenamefont
  {Chaudhuri}, \citenamefont {Berthier},\ and\ \citenamefont
  {Kob}}]{universal2007chaudhuri}%
  \BibitemOpen
  \bibfield  {author} {\bibinfo {author} {\bibfnamefont {P.}~\bibnamefont
  {Chaudhuri}}, \bibinfo {author} {\bibfnamefont {L.}~\bibnamefont
  {Berthier}},\ and\ \bibinfo {author} {\bibfnamefont {W.}~\bibnamefont
  {Kob}},\ }\bibfield  {title} {\bibinfo {title} {Universal nature of particle
  displacements close to glass and jamming transitions},\ }\href
  {https://doi.org/10.1103/PhysRevLett.99.060604} {\bibfield  {journal}
  {\bibinfo  {journal} {Phys. Rev. Lett.}\ }\textbf {\bibinfo {volume} {99}},\
  \bibinfo {pages} {060604} (\bibinfo {year} {2007})}\BibitemShut {NoStop}%
\bibitem [{\citenamefont {Palasantzas}\ and\ \citenamefont
  {Krim}(1993)}]{effect1993krim}%
  \BibitemOpen
  \bibfield  {author} {\bibinfo {author} {\bibfnamefont {G.}~\bibnamefont
  {Palasantzas}}\ and\ \bibinfo {author} {\bibfnamefont {J.}~\bibnamefont
  {Krim}},\ }\bibfield  {title} {\bibinfo {title} {Effect of the form of the
  height-height correlation function on diffuse x-ray scattering from a
  self-affine surface},\ }\href {https://doi.org/10.1103/PhysRevB.48.2873}
  {\bibfield  {journal} {\bibinfo  {journal} {Phys. Rev. B}\ }\textbf {\bibinfo
  {volume} {48}},\ \bibinfo {pages} {2873} (\bibinfo {year}
  {1993})}\BibitemShut {NoStop}%
\bibitem [{\citenamefont {Berthier}\ \emph {et~al.}(2004)\citenamefont
  {Berthier}, \citenamefont {Chandler},\ and\ \citenamefont
  {Garrahan}}]{LBerthier2005}%
  \BibitemOpen
  \bibfield  {author} {\bibinfo {author} {\bibfnamefont {L.}~\bibnamefont
  {Berthier}}, \bibinfo {author} {\bibfnamefont {D.}~\bibnamefont {Chandler}},\
  and\ \bibinfo {author} {\bibfnamefont {J.~P.}\ \bibnamefont {Garrahan}},\
  }\bibfield  {title} {\bibinfo {title} {Length scale for the onset of fickian
  diffusion in supercooled liquids},\ }\href
  {https://doi.org/10.1209/epl/i2004-10401-5} {\bibfield  {journal} {\bibinfo
  {journal} {Europhysics Letters}\ }\textbf {\bibinfo {volume} {69}},\ \bibinfo
  {pages} {320} (\bibinfo {year} {2004})}\BibitemShut {NoStop}%
\bibitem [{\citenamefont {Ramos}\ \emph {et~al.}(2011)\citenamefont {Ramos},
  \citenamefont {Oguni}, \citenamefont {Ishii},\ and\ \citenamefont
  {Nakayama}}]{ramos2011character}%
  \BibitemOpen
  \bibfield  {author} {\bibinfo {author} {\bibfnamefont {S.~L. L.~M.}\
  \bibnamefont {Ramos}}, \bibinfo {author} {\bibfnamefont {M.}~\bibnamefont
  {Oguni}}, \bibinfo {author} {\bibfnamefont {K.}~\bibnamefont {Ishii}},\ and\
  \bibinfo {author} {\bibfnamefont {H.}~\bibnamefont {Nakayama}},\ }\bibfield
  {title} {\bibinfo {title} {Character of devitrification, viewed from
  enthalpic paths, of the vapor-deposited ethylbenzene glasses},\ }\href
  {https://doi.org/10.1021/jp203612s} {\bibfield  {journal} {\bibinfo
  {journal} {The Journal of Physical Chemistry B}\ }\textbf {\bibinfo {volume}
  {115}},\ \bibinfo {pages} {14327} (\bibinfo {year} {2011})}\BibitemShut
  {NoStop}%
\bibitem [{\citenamefont {Beasley}\ \emph {et~al.}(2019)\citenamefont
  {Beasley}, \citenamefont {Bishop}, \citenamefont {Kasting},\ and\
  \citenamefont {Ediger}}]{beasley2019vapor}%
  \BibitemOpen
  \bibfield  {author} {\bibinfo {author} {\bibfnamefont {M.~S.}\ \bibnamefont
  {Beasley}}, \bibinfo {author} {\bibfnamefont {C.}~\bibnamefont {Bishop}},
  \bibinfo {author} {\bibfnamefont {B.~J.}\ \bibnamefont {Kasting}},\ and\
  \bibinfo {author} {\bibfnamefont {M.~D.}\ \bibnamefont {Ediger}},\ }\bibfield
   {title} {\bibinfo {title} {Vapor-deposited ethylbenzene glasses approach
  “ideal glass” density},\ }\href
  {https://doi.org/10.1021/acs.jpclett.9b01508} {\bibfield  {journal} {\bibinfo
   {journal} {The Journal of Physical Chemistry Letters}\ }\textbf {\bibinfo
  {volume} {10}},\ \bibinfo {pages} {4069} (\bibinfo {year}
  {2019})}\BibitemShut {NoStop}%
\bibitem [{\citenamefont {Jack}\ and\ \citenamefont
  {Berthier}(2016)}]{jack2016}%
  \BibitemOpen
  \bibfield  {author} {\bibinfo {author} {\bibfnamefont {R.~L.}\ \bibnamefont
  {Jack}}\ and\ \bibinfo {author} {\bibfnamefont {L.}~\bibnamefont
  {Berthier}},\ }\bibfield  {title} {\bibinfo {title} {The melting of stable
  glasses is governed by nucleation-and-growth dynamics},\ }\href
  {https://doi.org/10.1063/1.4954327} {\bibfield  {journal} {\bibinfo
  {journal} {The Journal of chemical physics}\ }\textbf {\bibinfo {volume}
  {144}},\ \bibinfo {pages} {244506} (\bibinfo {year} {2016})}\BibitemShut
  {NoStop}%
\bibitem [{\citenamefont {Guiselin}\ \emph
  {et~al.}(2022{\natexlab{b}})\citenamefont {Guiselin}, \citenamefont
  {Tarjus},\ and\ \citenamefont {Berthier}}]{guiselin2022glass}%
  \BibitemOpen
  \bibfield  {author} {\bibinfo {author} {\bibfnamefont {B.}~\bibnamefont
  {Guiselin}}, \bibinfo {author} {\bibfnamefont {G.}~\bibnamefont {Tarjus}},\
  and\ \bibinfo {author} {\bibfnamefont {L.}~\bibnamefont {Berthier}},\
  }\bibfield  {title} {\bibinfo {title} {Is glass a state of matter?},\ }\href
  {https://doi.org/10.13036/17533562.63.5.15} {\bibfield  {journal} {\bibinfo
  {journal} {Physics and Chemistry of Glasses-European Journal of Glass Science
  and Technology Part B}\ }\textbf {\bibinfo {volume} {63}},\ \bibinfo {pages}
  {136} (\bibinfo {year} {2022}{\natexlab{b}})}\BibitemShut {NoStop}%
\bibitem [{\citenamefont {Collins}\ \emph {et~al.}(2012)\citenamefont
  {Collins}, \citenamefont {Cochran}, \citenamefont {Yan}, \citenamefont
  {Gann}, \citenamefont {Hub}, \citenamefont {Fink}, \citenamefont {Wang},
  \citenamefont {Schuettfort}, \citenamefont {McNeill}, \citenamefont
  {Chabinyc} \emph {et~al.}}]{collins2012polarized}%
  \BibitemOpen
  \bibfield  {author} {\bibinfo {author} {\bibfnamefont {B.~A.}\ \bibnamefont
  {Collins}}, \bibinfo {author} {\bibfnamefont {J.~E.}\ \bibnamefont
  {Cochran}}, \bibinfo {author} {\bibfnamefont {H.}~\bibnamefont {Yan}},
  \bibinfo {author} {\bibfnamefont {E.}~\bibnamefont {Gann}}, \bibinfo {author}
  {\bibfnamefont {C.}~\bibnamefont {Hub}}, \bibinfo {author} {\bibfnamefont
  {R.}~\bibnamefont {Fink}}, \bibinfo {author} {\bibfnamefont {C.}~\bibnamefont
  {Wang}}, \bibinfo {author} {\bibfnamefont {T.}~\bibnamefont {Schuettfort}},
  \bibinfo {author} {\bibfnamefont {C.~R.}\ \bibnamefont {McNeill}}, \bibinfo
  {author} {\bibfnamefont {M.~L.}\ \bibnamefont {Chabinyc}}, \emph {et~al.},\
  }\bibfield  {title} {\bibinfo {title} {Polarized x-ray scattering reveals
  non-crystalline orientational ordering in organic films},\ }\href
  {https://doi.org/10.1038/nmat3310} {\bibfield  {journal} {\bibinfo  {journal}
  {Nature materials}\ }\textbf {\bibinfo {volume} {11}},\ \bibinfo {pages}
  {536} (\bibinfo {year} {2012})}\BibitemShut {NoStop}%
\bibitem [{\citenamefont {Ferron}\ \emph {et~al.}(2022)\citenamefont {Ferron},
  \citenamefont {Thelen}, \citenamefont {Bagchi}, \citenamefont {Deng},
  \citenamefont {Gann}, \citenamefont {de~Pablo}, \citenamefont {Ediger},
  \citenamefont {Sunday},\ and\ \citenamefont
  {DeLongchamp}}]{ferron2022characterization}%
  \BibitemOpen
  \bibfield  {author} {\bibinfo {author} {\bibfnamefont {T.~J.}\ \bibnamefont
  {Ferron}}, \bibinfo {author} {\bibfnamefont {J.~L.}\ \bibnamefont {Thelen}},
  \bibinfo {author} {\bibfnamefont {K.}~\bibnamefont {Bagchi}}, \bibinfo
  {author} {\bibfnamefont {C.}~\bibnamefont {Deng}}, \bibinfo {author}
  {\bibfnamefont {E.}~\bibnamefont {Gann}}, \bibinfo {author} {\bibfnamefont
  {J.~J.}\ \bibnamefont {de~Pablo}}, \bibinfo {author} {\bibfnamefont
  {M.}~\bibnamefont {Ediger}}, \bibinfo {author} {\bibfnamefont {D.~F.}\
  \bibnamefont {Sunday}},\ and\ \bibinfo {author} {\bibfnamefont {D.~M.}\
  \bibnamefont {DeLongchamp}},\ }\bibfield  {title} {\bibinfo {title}
  {Characterization of the interfacial orientation and molecular conformation
  in a glass-forming organic semiconductor},\ }\href
  {https://doi.org/10.1021/acsami.1c19948} {\bibfield  {journal} {\bibinfo
  {journal} {ACS Applied Materials \& Interfaces}\ }\textbf {\bibinfo {volume}
  {14}},\ \bibinfo {pages} {3455} (\bibinfo {year} {2022})}\BibitemShut
  {NoStop}%
\bibitem [{\citenamefont {Freychet}\ \emph {et~al.}(2021)\citenamefont
  {Freychet}, \citenamefont {Gann}, \citenamefont {Thomsen}, \citenamefont
  {Jiao},\ and\ \citenamefont {McNeill}}]{freychet2021resonant}%
  \BibitemOpen
  \bibfield  {author} {\bibinfo {author} {\bibfnamefont {G.}~\bibnamefont
  {Freychet}}, \bibinfo {author} {\bibfnamefont {E.}~\bibnamefont {Gann}},
  \bibinfo {author} {\bibfnamefont {L.}~\bibnamefont {Thomsen}}, \bibinfo
  {author} {\bibfnamefont {X.}~\bibnamefont {Jiao}},\ and\ \bibinfo {author}
  {\bibfnamefont {C.~R.}\ \bibnamefont {McNeill}},\ }\bibfield  {title}
  {\bibinfo {title} {Resonant tender x-ray diffraction for disclosing the
  molecular packing of paracrystalline conjugated polymer films},\ }\href
  {https://doi.org/10.1021/jacs.0c10721} {\bibfield  {journal} {\bibinfo
  {journal} {Journal of the American Chemical Society}\ }\textbf {\bibinfo
  {volume} {143}},\ \bibinfo {pages} {1409} (\bibinfo {year}
  {2021})}\BibitemShut {NoStop}%
\bibitem [{\citenamefont {de~Saint-Exup\'ery}(1943)}]{prince}%
  \BibitemOpen
  \bibfield  {author} {\bibinfo {author} {\bibfnamefont {A.}~\bibnamefont
  {de~Saint-Exup\'ery}},\ }\href@noop {} {\emph {\bibinfo {title} {The little
  prince}}}\ (\bibinfo  {publisher} {Reynal and Hitchcock, New York},\ \bibinfo
  {year} {1943})\BibitemShut {NoStop}%
\end{thebibliography}%

\end{document}